\documentclass[preprint,11pt]{elsarticle}

\usepackage[a4paper,top=1.5cm,bottom=1.5cm,left=2cm,right=2cm]{geometry}
\usepackage{algorithm}
\usepackage{algpseudocode}
\usepackage{amsthm}
\usepackage{amsmath}
\usepackage{amssymb}
\usepackage{amsfonts}
\usepackage{bbm}
\usepackage{graphicx}
\usepackage{epstopdf}
\usepackage{dcolumn}
\usepackage{bm}
\usepackage{braket}
\usepackage{color}
\usepackage{enumerate}
\usepackage{mathtools}
\usepackage{makecell}
\usepackage{tikz-cd}
\usepackage{mathtools}
\usepackage{amssymb}
\usepackage{enumitem}
\usepackage{bm} 
\usepackage{multirow}
\usepackage{booktabs}
\usepackage{array}
\usepackage[colorlinks=true, allcolors=blue]{hyperref}

\usepackage{tikz}
\usepackage{booktabs}
\usepackage{multirow}
\usepackage{makecell}

\usepackage{tabularx}
\usepackage{float} 

\theoremstyle{plain}\newtheorem{theorem}{Theorem}
\theoremstyle{definition}\newtheorem{remark}{Remark}
\theoremstyle{definition}
\theoremstyle{plain}\newtheorem{Pp}[theorem]{Proposition}
\theoremstyle{plain}
\theoremstyle{plain}\newtheorem{lemma}{Lemma}
\theoremstyle{plain}
\newtheorem{defi}{Definition}
\newtheorem{thm}{Theorem}

\allowdisplaybreaks

\usepackage{amssymb}
\usepackage{amsmath}

\journal{Arxiv}

\begin{document}

\begin{frontmatter}



\title{Revisiting finite Abelian hidden subgroup problem and its distributed exact quantum algorithm} 


\author{Ziyuan Dong\fnref{label1,label2}}
\author{Xiang Fan\fnref{label3}} 
\author{Tengxun Zhong\fnref{label3}}
\author{Daowen Qiu\corref{mycorrespondingauthor}\fnref{label1,label2}}

\affiliation[label1]{organization={Institute of Quantum Computing and Software, School of Computer Science and Engineering, Sun Yat-sen University},
            city={Guangzhou},
           postcode={510006}, 
           country={China}}

\affiliation[label2]{organization={The Guangdong Key Laboratory of Information Security Technology, Sun Yat-sen University},
          city={Guangzhou},
           postcode={510006},
          country={China}}

\affiliation[label3]{organization={School of Mathematics, Sun Yat-sen University},
            city={Guangzhou}, 
            postcode={510275},
            country={China}}

\cortext[mycorrespondingauthor]{Corresponding author at: School of Computer Science and Engineering, Sun Yat-sen University, Guangzhou 510006, China.
\\
E-mail addresses:
\href{mailto:dongzy5@mail2.sysu.edu.cn}{dongzy5@mail2.sysu.edu.cn} (Z. Dong),
\href{mailto:issqdw@mail.sysu.edu.cn}{issqdw@mail.sysu.edu.cn} (D. Qiu).
}

\begin{abstract}
We revisit the finite Abelian hidden subgroup problem (AHSP) from a mathematical perspective and make the following contributions. First, by employing amplitude amplification, we present an exact quantum algorithm for the finite AHSP, our algorithm is more concise than the previous exact algorithm and applies to any finite Abelian group. Second, utilizing the Chinese Remainder Theorem, we propose a distributed exact quantum algorithm for finite AHSP, which requires fewer qudits, lower quantum query complexity, and no quantum communication. We further show that our distributed approach can be extended to certain classes of non-Abelian groups. Finally, we develop a parallel exact classical algorithm for finite AHSP with reduced query complexity; even without parallel execution, the total number of queries across all nodes does not exceed that of the original centralized algorithm under mild conditions.

\end{abstract}

\begin{keyword}
Abelian hidden subgroup problem, exact quantum algorithm, distributed quantum algorithms, LOCC protocol
\end{keyword}

\end{frontmatter}


\section{Introduction}{\label{Sec1}}

Hidden subgroup problem (HSP) takes as input a group $G$, a finite set $S$, and a black-box function $f: G \to S$. It is promised that there exists a subgroup $K \leq G$ such that $f(x) = f(y)$ if and only if $xK = yK$. The goal of HSP is to identify the subgroup \( K \). HSP~\cite{kitaev1995quantum} is a central problem in quantum computing, which encompasses most of the known exponential speedup applications of quantum Fourier transform~\cite{nielsen_quantum_2010}. There are many concrete instances for HSP, such as order finding problem~\cite{shor_polynomial-time_1997}, discrete logarithm problem~\cite{shor_polynomial-time_1997}, Simon's problem~\cite{simon_power_1997}, and generalized Simon's problem~\cite{kaye_introduction_2007}. 

For finite Abelian hidden subgroup problem (AHSP), researchers have developed a relatively mature research framework, in which quantum Fourier transform is a key technique. For achieving success probability $1-\epsilon$, the standard quantum algorithm for finite AHSP requires $\lfloor 4/\epsilon\rfloor\mathrm{rank}(G)$ quantum queries~\cite{koiran2007quantum} or $\lceil \log_2 |G| + \log_2 (1/\epsilon) + 2 \rceil$ quantum queries~\cite{lomont2004hidden}, providing an exponential speedup over classical algorithms. Research of HSP on non-Abelian groups is more challenging and remains an active field, where the Fourier transform, although well-defined, may not have an efficient quantum implementation~\cite{lomont2004hidden}. For the dihedral group $G = D_N$ (of order $2N$), Kuperberg proposed two subexponential-time quantum algorithms for the HSP over $G$~\cite{kuperberg2005subexponential,kuperberg2013another}, both with time and query complexity $2^{\mathcal{O}(\sqrt{\log N})}$. For the general HSP, the universal quantum algorithm in~\cite{ettinger2004quantum} achieves an exponential advantage in query complexity over any classical counterpart, requiring only $\mathcal{O}((\log|G|)^2 + \log(1/\epsilon))$ queries to ensure success with probability $1-\epsilon$; nevertheless, its time complexity remains exponential, thus limiting its practical applicability.

For the classical query complexity of HSP, deterministic algorithms have been established for various cases.
Nayak's deterministic algorithm~\cite{nayak2022deterministic} requires $\mathcal{O}\left(\log |K| \cdot \sqrt{\frac{|G|}{|K|}}\right)$ queries for AHSP and $\mathcal{O}\left(\log |K| \cdot \sqrt{\frac{|G|}{|K|} \log \frac{|G|}{|K|}}\right)$  
queries for a broad class of non-Abelian groups. Meanwhile, the deterministic algorithm in~\cite{ye2022deterministic} finds the subgroup $K$ with $\mathcal{O}\left(\sqrt{\frac{|G|}{|K|}\log|K|} + \log|K|\right)$ queries.

While quantum algorithms show strong theoretical performance, putting them into practice is hard because of current hardware limits. It is difficult to simultaneously achieve a large number of qubits and deep quantum circuits, both of which are essential for universal quantum computing. 
In the Noisy Intermediate-Scale Quantum (NISQ) era~\cite{preskill_quantum_2018}, one promising path forward is distributed quantum computing, which refers to a computational model that achieves large-scale data processing tasks by aggregating quantum computer resources from various nodes. Compared with centralized quantum computing, distributed quantum computing may involve a reduction in the number of qubits and circuit depth, while also achieving noise resistance as an added benefit.

Recently, there have been many works on distributed quantum algorithms~\cite{goos_distributed_2003,beals_efficient_2013,caleffi_quantum_2018}. In 2017, Qiu et al.~\cite{Qiu2017DQC} proposed a distributed quantum phase estimation algorithm. In 2018, Le Gall et al.~\cite{le2018sublinear} studied distributed quantum algorithms in the quantum CONGEST model. The following year, Izumi et al.~\cite{izumi2019quantum} investigated quantum distributed algorithms in the CONGEST-CLIQUE model. In 2022, Anshu et al.~\cite{anshu2022distributed} introduced a distributed quantum algorithm for inner product estimation, while Tan et al.~\cite{Tan2022DQCSimon} developed a distributed quantum algorithm for Simon's problem. In 2023, Xiao et al.~\cite{Xiao2023DQAShor} proposed a distributed Shor's algorithm. In 2024, Qiu et al.~\cite{Qiu22} proposed  a distributed Grover's algorithm; Li et al.~\cite{li2024exact} devised a distributed quantum algorithm for generalized Simon’s problem; and Coiteux-Roy et al.~\cite{coiteux2024no} established that there is no distributed quantum advantage for approximate graph coloring. In 2025, Qiu et al.~\cite{qiu2025error} proposed a universal error correction scheme for distributed quantum computing.

Among the aforementioned works, the distributed algorithms for Simon's problem, the generalized Simon's problem, and Shor's algorithm all represent specific instances of distributed AHSP algorithms. Furthermore, all existing distributed quantum algorithms require quantum communication.

In this paper, we first present the iteration count for achieving success probability $1-\epsilon$ in the standard quantum algorithm for finite AHSP to be either $\mathrm{rank}(G) + \lceil \log_2 (2/\epsilon) \rceil$ or $\mathrm{len}(G) + \lceil \log_2 (1/\epsilon) \rceil$. The former offers an exponential improvement in $\epsilon$-dependence over the prior bound $\lfloor 4/\epsilon \rfloor \mathrm{rank}(G)$~\cite{koiran2007quantum}, while the latter improves upon $\lceil \log_2 |G| + \log_2 (1/\epsilon) + 2 \rceil$~\cite{lomont2004hidden}.

Employing amplitude amplification, we present an exact quantum algorithm for finite AHSP, provided the order $|K|$ of hidden subgroup $K$ is known in advance. Our algorithm is more concise than the previous exact algorithm~\cite{imran2022exact} for finite AHSP.  The method in~\cite{imran2022exact} is only applicable to groups of the form $G=\bigl(\mathbb{Z}_{m^k}\bigr)^n$ and requires $\mathcal{O}\big(nk \cdot \log^{2} m\big)$ quantum queries, equivalent to $\mathcal{O}\big(\mathrm{len}(G) \log^{2} m\big)$. In contrast, Algorithm~\ref{algorithm3} applies to any finite Abelian group $G$ and achieves a significantly lower complexity of $3(\mathrm{len}(G) - \mathrm{len}(K))$. This complexity is independent of $m$ and is further reduced by incorporating the term $\mathrm{len}(K)$.

After that, using the Chinese Remainder Theorem, we propose a distributed exact quantum algorithm for finite AHSP (Algorithm~\ref{EDK}), which is a pure parallel algorithm and needs no quantum communication. Specifically, consider the finite AHSP with domain (group) $G =\mathop{\oplus}\limits_{i=1}^m G_i=\mathop{\oplus}\limits_{i=1}^m \mathop{\oplus}\limits_{j = 1}^{r_i} \mathbb{Z}_{p_i^{\alpha_{ij}}}$, where $p_1,\dots,p_m$ are distinct primes and $\alpha_{ij} \ge 1$, codomain (finite set) $S$, and hidden subgroup $K$. Algorithm~\ref{EDK} employs $m$ nodes and significantly reduces the resource requirements per node. In terms of qudit number, each node $i$ ($1 \leq i \leq m$) requires only $N(G_i) + N(S)$ qudits, much lower than that of $N(G) + N(S)$ qudits in the original centralized algorithms (Algorithms~\ref{algorithm1} and~\ref{algorithm3}). Here, $N(\cdot)$ denotes the number of qudits required to encode the corresponding group $G$ or $G_i$, and set $S$. For quantum queries, the per‑node complexity of Algorithm~\ref{EDK} is $\mathop{\max}\limits_{1 \leq i \leq m} 3 \left( \mathrm{len}(G_i) - \mathrm{len}(K_i) \right)$, much lower than the centralized complexity $3\bigl(\operatorname{len}(G) - \operatorname{len}(K)\bigr)$ of Algorithm~\ref{algorithm3}. We compare distributed quantum Algorithm~\ref{EDK} against centralized quantum algorithms in Table~\ref{fenbushiliangzisuanfabiao}. By reducing qudits per node, our distributed quantum Algorithm~\ref{EDK} needs shallower oracle circuits, enhancing noise resistance for NISQ-era implementation.

\begin{table}[htbp]
\footnotesize
\setlength{\tabcolsep}{4pt}
\caption{Comparison of distributed quantum Algorithm~\ref{EDK} with other algorithms}
\label{fenbushiliangzisuanfabiao}
\centering
\begin{tabular}{@{}lcccc@{}}
\toprule
\quad\textbf{Type} & \textbf{Algorithms} & \makecell[c]{\textbf{Number of qudits}\\\textbf{at each node}} & \textbf{Query complexity} & \makecell[c]{\textbf{Success}\\\textbf{probability}} \\
\midrule

\multirow{3}{*}{Centralized} 
& Algorithm~\ref{algorithm1} & $N(G) + N(S)$ & 
$\begin{array}{l}
\min \left\{ \mathrm{rank}(G)+\lceil\log_2(2/\epsilon)\rceil, \ \mathrm{len}(G)+\lceil\log_2(1/\epsilon)\rceil \right\}
\end{array}$ 
& $\geq 1-\epsilon$ \\

\addlinespace

& Algorithm~\ref{algorithm3} & $N(G) + N(S)$ & 
$3(\mathrm{len}(G)-\mathrm{len}(K))$ 
& 100\% (exact)\\

Distributed
& Algorithm~\ref{EDK} & $N(G_i) + N(S)$ & 
$\max\limits_{1 \leq i \leq m} 3 \left( \mathrm{len}(G_i) - \mathrm{len}(K_i) \right)$ 
& 100\% (exact)\\
\bottomrule
\end{tabular}

\vspace{6pt}
\footnotesize{Note: $N(\cdot)$ denotes the number of qudits to encode the corresponding group $G$ or $G_i$, and set $S$. $\mathrm{len}(\cdot)$ denotes chain length, $\mathrm{rank}(\cdot)$ is the minimal number of generators.}
\end{table}

We also develop a  classical parallel exact algorithm (EDCK) for finite AHSP (Algorithm~\ref{EDCK}), which also needs $m$ nodes and lower query complexity. Specially, the classical query complexity of our parallel Algorithm~\ref{EDCK} is only $\mathcal{O}\left(\mathop{\max}\limits_{1\leq i\leq m}(\sqrt{\dfrac{|G_i|}{|K_i|}\log|K_i|}+\log|K_i|)\right)$, much lower than that of $\mathcal{O}(\sqrt{\dfrac{|G|}{|K|}\log|K|}+\log|K|)$ in the original classical exact algorithm in~\cite{ye2022deterministic}. Even without parallel execution, the total classical query complexity satisfies
    \[
\mathcal{O}\left( \sum_{i=1}^{m} \left( \sqrt{\frac{|G_i|}{|K_i|} \log |K_i|} + \log |K_i| \right) \right) \subseteq \mathcal{O}\left( \sqrt{\frac{|G|}{|K|} \cdot \log |K|} + \log |K| \right).
\]
    under the sufficient condition (not necessary) $K_i \neq G_i$ ($i = 1, 2, \ldots, m$). This demonstrates that our classical parallel algorithm achieves an \emph{inherent speedup}. We compare parallel exact classical parallel Algorithm~\ref{EDCK} against state-of-the-art classical deterministic algorithms in Table~\ref{fenbushisuanfabiao}.

\begin{table}[htbp]
\footnotesize
\setlength{\tabcolsep}{4pt}
\caption{Comparison of parallel classical Algorithm~\ref{EDCK} with other algorithms}
\label{fenbushisuanfabiao}
\centering
\begin{tabular}{@{}lcccc@{}}
\toprule
\quad\textbf{Type} & \textbf{Algorithms} & 
\textbf{Query complexity}& 
\makecell[c]{\textbf{Success}\\\textbf{probability}}\\
\midrule

\multirow{5}{*}{Centralized}
& {\makecell[c]{Classical deterministic\\ algorithm for finite\\AHSP  in~\cite{nayak2022deterministic}}} 
& $\mathcal{O}\left(\sqrt{\dfrac{|G|}{|K|}}\cdot\log |K|\right)$ 
& 100\% (exact)\\

\addlinespace

&\makecell[c]{Classical deterministic\\algorithm for finite\\ AHSP in~\cite{ye2022deterministic}} 
& $\mathcal{O}\left(\sqrt{\dfrac{|G|}{|K|}\log|K|}+\log|K|\right)$ 
& 100\% (exact) \\

\quad Parallel
& Algorithm~\ref{EDCK} 
& $\mathcal{O}\left(\max\limits_{1\leq i\leq m}\left(\sqrt{\dfrac{|G_i|}{|K_i|}\log|K_i|}+\log|K_i|\right)\right)$ 
& 100\% (exact)\\
\bottomrule
\end{tabular}
\end{table}

Our Algorithm~\ref{EDK} is the first distributed quantum algorithm for finite AHSP without quantum communication. As shown in Table~\ref{fenbushisimonbijiaobiao}, it offers two fundamental advances over prior distributed algorithms for Simon's problem~\cite{Tan2022DQCSimon} and generalized Simon's problem~\cite{li2024exact}. First, Algorithm~\ref{EDK} is the only pure LOCC (Local Operations and Classical Communication) algorithm eliminating the need for quantum communication, whereas algorithms in~\cite{Tan2022DQCSimon, li2024exact} require $\mathcal{O}((n-t)(2^{t} - 1)(n-t+l))$ quantum communication. Second, Algorithm~\ref{EDK}'s node count ($m$) is determined by the number of prime factors of $|G|$, reducing from $2^t$ (exponential) nodes in prior works~\cite{Tan2022DQCSimon, li2024exact}, which marks a substantial improvement in resource efficiency.

\begin{table}[htbp]
\footnotesize
\setlength{\tabcolsep}{4pt}
\centering  
\caption{Comparison of Algorithm~\ref{EDK} with distributed quantum algorithms for Simon's and generalized Simon's problem}  
\label{fenbushisimonbijiaobiao}
\begin{tabular}{@{}lccccc@{}}   
\toprule
\quad\textbf{Algorithms} & \textbf{Group \(G\)} & \makecell[c]{\textbf{Success}\\\textbf{probability}} & \textbf{Applicable Problems} & \textbf{Nodes} & \makecell[c]{\textbf{Quantum}\\\textbf{communication}\\\textbf{complexity}} \\   
\midrule
\quad Algorithm~\ref{EDK} & $G=\mathop{\oplus}\limits_{i=1}^m \mathop{\oplus}\limits_{j = 1}^{r_i} \mathbb{Z}_{p_i^{\alpha_{ij}}}$ & 100\% (exact) &finite  AHSP & $m$ & $0$ \\ 

\quad Algorithm~\ref{EDCK} & $G=\mathop{\oplus}\limits_{i=1}^m \mathop{\oplus}\limits_{j = 1}^{r_i} \mathbb{Z}_{p_i^{\alpha_{ij}}}$ & 100\% (exact) &finite AHSP & $m$ & $0$ \\ 

Algorithm in~\cite{Tan2022DQCSimon} & $G=(\mathbb{Z}_2)^n$ & $< 100\%$ (inexact) & Simon's problem & $2^t$ & $\mathcal{O}\big((n-t)(2^{t} - 1)(n-t + l)\big)$ \\  

Algorithm in~\cite{li2024exact} & $G=(\mathbb{Z}_2)^n$ & 100\% (exact) & generalized Simon's problem & $2^t$ & $\mathcal{O}\big((n-t)(2^{t} - 1)(n-t+l)\big)$ \\  
\bottomrule
\end{tabular} 
\vspace{6pt}

\footnotesize 
Note: $t$ denotes the number of bits by which the input of function $f$ is decomposed; $l$ denotes the number of output bits of function $f$.
\end{table}

The remainder of this paper is organized as follows. In Sec.~\ref{Sec2}, we recall the definition of HSP and AHSP. Then we give the properties of $\mathrm{rank}(G)$ and $\mathrm{len}(G)$. In Sec.~\ref{Sec3}, we revisit the standard quantum algorithm for finite AHSP. We then proceed to determine, from the dual perspectives of group rank and chain length, the number of iterations needed to guarantee a success probability of \(1 - \epsilon\). Using the quantum amplitude amplification subroutine, we give the exact quantum algorithm for finite AHSP in Sec.~\ref{Sec4}. In Sec.~\ref{Sec5}, we first present two theorems serving as the foundation of our distributed quantum algorithm. Then we give an exact distributed quantum algorithm for finite AHSP and prove its correctness. Furthermore, we show that our distributed method can be generalized to certain classes of finite non-Abelian groups. In Sec.~\ref{Sec6}, we present an exact classical parallel algorithm for finite AHSP and prove its key advantage. Finally, in Sec.~\ref{Sec7}, we summarize the main conclusions and highlight some problems deserving further investigation.

\section{Preliminaries}\label{Sec2}

In this section, we first recall HSP and AHSP, followed by the definition of orthogonal subgroup \( K^\perp \). Then, we establish key properties of \(\mathrm{rank}(G)\) and \(\mathrm{len}(G)\), which are crucial for analyzing the quantum query complexity of AHSP.

\subsection{Overview of hidden subgroup problem}

\begin{defi}[\textbf{Hidden Subgroup Problem (HSP)}\cite{kitaev1995quantum, nielsen_quantum_2010}]\label{HSPdingyi}
Let \( G \) be a group and \( S \) be a finite set. Given a black-box function \( f: G \to S \), suppose there exists an unknown subgroup \( K \leq G \) such that for all \( x, y \in G \),
\begin{align*}
    f(x) = f(y) \quad \text{if and only if} \quad xK = yK,\\
\text{i.e., } x^{-1}y\in K\text{ or } y^{-1}x \in K.
\end{align*}
The goal of HSP is to identify subgroup \( K \) (or a generating set for \( K \)) by querying \( f \).
\end{defi}

\begin{remark}
$xK$ and $yK$ are both left coset. The function \( f \) is \textit{constant} on left cosets of \( K \) and \textit{distinct} across different cosets. When $G$ is a finite Abelian group, the problem is specifically referred to as the finite Abelian Hidden Subgroup Problem (AHSP).

\end{remark}

Suppose we have an oracle that can query the value of function $f$. For any $g\in G$ and $b\in S$, if we input $|g\rangle|b\rangle$ into the oracle, then $|g\rangle|b+f(g)\rangle$ is obtained.

Any finite Abelian group $G$ is isomorphic to a direct sum of cyclic groups of prime-power order:
\begin{equation} \label{elementary factors}
G \cong \bigoplus_{i=1}^k \mathbb{Z}/p_i^{\alpha_i}\mathbb{Z}. 
\end{equation}
Letting $N_i = p_i^{\alpha_i}$ for brevity, the decomposition satisfies the following conditions:
\begin{itemize}
    \item The primes $p_i$ are not necessarily distinct.
    \item $\alpha_i$ is a positive integer.
\end{itemize}

In this decomposition, every element $x \in G$ is represented as $k$-tuples $x = (x_1, \dots, x_k)$, where $x_i \in \{0, 1, \dots, N_i-1\}$ for $1 \leq i \leq k$. And we have the definition of bilinear form.

\begin{defi}[\textbf{Bilinear Form}]\label{shuangxianxingxingshi}
For any $x, y \in G$, define a bilinear map $\langle\cdot,\cdot\rangle : G \times G \to \mathbb{R}/\mathbb{Z}$ by
\begin{equation*}
\langle x, y\rangle := \sum_{i=1}^{k} \frac{x_i \cdot y_i}{N_i}\pmod 1,
\end{equation*}
where the result is taken in $[0,1)$ for computational convenience.
\end{defi}

Then, we have the definition of orthogonal subgroup $K^{\perp}$.

\begin{defi}[\cite{kaye_introduction_2007}]\label{zhengjiaoqun}
Let $G$ be a finite Abelian group. For any subgroup $K \leq G$, the subgroup $K^{\perp}$ is defined as:
\begin{equation*}
K^{\perp} := \left\{ g \in G \mid \langle k, g \rangle = 0 \pmod 1 \text{ for all } k \in K \right\}.
\end{equation*}
\end{defi}

\begin{lemma}[\cite{serre1977linear, lomont2004hidden}]\label{zhengjiaotonggoudingli}
Let $G$ be a finite Abelian group and $K \leq G$ a subgroup. Then:

\begin{enumerate}[label=\textup{(\roman*)}, itemsep=0pt,align=left,labelwidth=1.5em]
   \item $K^{\perp}\cong G/K$;
    \item $K^{\perp\perp}:=(K^{\perp})^{\perp}=K$.
\end{enumerate}

\end{lemma}
\subsection{Properties of $\mathrm{rank}(G)$ and $\mathrm{len}(G)$}

$\mathrm{rank}(G)$ is defined as the  minimal cardinality of a generating set of $G$. And \textbf{chain length}  $\mathrm{len}(G)$ is defined as follows:

\begin{defi}[\cite{hungerford2012algebra}]\label{lianchangdingyi}
A \textbf{composition series} of a group $G$ is a maximal subgroup chain
\[
\{e\} = G_0 \subsetneqq G_1 \subsetneqq \cdots \subsetneqq G_r = G
\]
where each $G_{i-1} \trianglelefteq G_i$ and $G_i/G_{i-1}$ is simple. The integer $r$ is called the \textbf{chain length} of $G$, denoted r=$\mathrm{len}(G)$.
\end{defi}

\begin{remark}
For a finite group $G$, its \textbf{chain length} $\mathrm{len}(G)$ is a well-defined invariant, even though its \textbf{composition series} are not necessarily unique. If a finite Abelian group $G$ has order $|G| = \prod\limits_{i=1}^{m} p_i^{e_i}$, then its chain length is $\mathrm{len}(G) = \sum\limits_{i=1}^{m} e_i$.
\end{remark}

Proposition~\ref{zhihelianchangguanxidingli} demonstrates that for the groups $G=(\mathbb{Z}_p)^k$ which underlie Simon's problem and generalized Simon's problem, the notions of rank and length coincide and are equivalent to the concept of dimension.

\begin{Pp}[\cite{dong2025probabilistic}]\label{zhihelianchangguanxidingli}
Let $G$ be a finite solvable group. Then $\mathrm{len}(G) \geq \mathrm{rank}(G)$. Moreover, if $G \cong (\mathbb{Z}_p)^k$ for some prime $p$ and integer $k$, then $\mathrm{len}(G) = \mathrm{rank}(G)$.
\end{Pp}

Proposition~\ref{lianchangguanxi} plays a crucial role in our analysis of the AHSP in Algorithms~\ref{algorithm1} and~\ref{algorithm3}. The additivity of chain length allows us to express $\mathrm{len}(K^{\perp})$ in terms of $\mathrm{len}(G)$ and $\mathrm{len}(K)$. This enables
us to determine query complexity directly from $\mathrm{len}(K^{\perp})$ rather than from $\mathrm{len}(G)$. Furthermore, Proposition~\ref{zhiguanxi} demonstrates that \textit{a priori} knowledge of $\mathrm{rank}(K)$ does not reduce the minimal number of algorithm iterations. Proposition~\ref{zhihuolianchangdingli} provides the foundation for determining the algorithm's complexity.

\begin{Pp}[\textbf{Additivity of Chain Length}~\cite{hungerford2012algebra}]\label{lianchangguanxi}
For a finite Abelian group $G$ and subgroup $K \subseteq G$:
\begin{equation*}
\mathrm{len}(G) = \mathrm{len}(K) + \mathrm{len}(K^{\perp}).
\end{equation*}
\end{Pp}
\begin{Pp}[\textbf{Subadditivity of Rank}~\cite{dong2025probabilistic}]\label{zhiguanxi}
For a finite Abelian group $G$ and subgroup $K \subseteq G$:
\begin{equation*}
\mathrm{rank}(G)\le \mathrm{rank}(K)+\mathrm{rank}(K^{\perp}).
\end{equation*}
\end{Pp}

\begin{Pp}[\textbf{Bounds for Generating Finite Nilpotent Groups}~\cite{dong2025probabilistic}]\label{zhihuolianchangdingli}

For any finite nilpotent group $G$ and $0<\epsilon<1$. Let $\varphi_k(G)$ denote the probability that $k$ elements $\{g_1,\ldots,g_k\}$, sampled uniformly and independently with replacement from $G$, generate the entire group $G$.

Then the following bounds hold:
\begin{enumerate}[label=\textup{(\roman*)}, itemsep=0pt,align=left,labelwidth=1.5em]
\item $\varphi_k(G) \ge 1-\epsilon$, if $k \ge \operatorname{rank}(G) + \left\lceil \log_2 \dfrac{2}{\epsilon} \right\rceil$.
\item $\varphi_k(G) \ge 1-\epsilon$, if $k \ge \operatorname{len}(G) + \left\lceil \log_2 \dfrac{1}{\epsilon} \right\rceil$.
\end{enumerate}
\end{Pp}

\section{Revisiting standard quantum algorithm for finite AHSP}\label{Sec3}

In Definition~\ref{HSPdingyi}, where the sole constraint on set $S$ is $|S| \ge |f(G)|$. We set $S = G$ for simplicity, without loss of generality. This choice covers the worst-case scenario $|K|=1$, which requires $|S| \ge |G|$, and allows us to use the same qudit representation for both the input and output registers of the quantum oracle, thereby simplifying resource analysis of qudit number of our algorithm.

We present the standard quantum algorithm~\cite{kaye_introduction_2007, qiu2025theoretical} for finite AHSP in Algorithm~\ref{algorithm1}, with the only modification in line 2. According to Theorem~\ref{suanfachaxunfuzadu}, the iteration count $h$ is set to the minimum of two values: $\min \left\{ \mathrm{rank}(G)+\lceil\log_2(2/\epsilon)\rceil, \ \mathrm{len}(G)+\lceil\log_2(1/\epsilon)\rceil \right\}$. If the chain length of the subgroup $K$ is known in advance—for instance, as in the generalized Simon's problem where the dimension $\dim(K)$ of $K$ is given—we may replace the expression $\mathrm{len}(G) + \lceil\log_2(1/\epsilon)\rceil$ with $\mathrm{len}(G) - \mathrm{len}(K) + \lceil\log_2(1/\epsilon)\rceil$.

\begin{algorithm}[h]
  \caption{Quantum algorithm for finite AHSP}\label{algorithm1}
  \begin{algorithmic}[1]
    \Statex \textbf{Input:} Integer $N_1,N_2,\cdots,N_k$, oracle $U_f$, success probability $1-\epsilon$
    \Statex \textbf{Output:} Subgroup $A$ containing hidden subgroup $K$
    \State $T_0 \gets \emptyset$; 
    \State $h\gets\min \left\{ \mathrm{rank}(G)+\lceil\log_2(2/\epsilon)\rceil, \ \mathrm{len}(G)+\lceil\log_2(1/\epsilon)\rceil \right\}$; \Comment{By Theorem~\ref{suanfachaxunfuzadu}}
    
    \For{$i = 1$ to $h$}
        \State $|\psi_0\rangle=|0\rangle|0\rangle \cdots|0\rangle|0_G\rangle\in\mathcal{H}_{N_1} \otimes \mathcal{H}_{N_2} \otimes \cdots \otimes \mathcal{H}_{N_k} \otimes \mathcal{H}_{G}$;
        \State $|\psi_1\rangle=({\rm QFT}_{N_1}\otimes{\rm QFT}_{N_2}\otimes\cdots\otimes{\rm QFT}_{N_k}\otimes \mathrm{I}_{G})|\psi_0\rangle$;
        \State $|\psi_2\rangle=U_f|\psi_1\rangle$;
        \State $|\psi_3\rangle=({\rm QFT}_{N_1}^{\dagger}\otimes{\rm QFT}_{N_2}^{\dagger}\otimes\cdots\otimes{\rm QFT}_{N_k}^{\dagger}\otimes \mathrm{I}_{G})|\psi_2\rangle$;
        \State Measure the first register to obtain $\bm{t}_i=(t_{i1}, t_{i2},\cdots, t_{ik})\in K^{\perp}$;
        \State Update $T_i\gets T_{i-1}\cup \{\bm{t}_i\}$.
    \EndFor
    
    \State Solve linear congruence system:
    \[
    \bm{W}\bm{x}^{\mathrm{T}} \equiv \bm{0}^{\mathrm{T}}  \pmod{1} \quad \text{where } \bm{W}= \begin{pmatrix} 
    t_{11}/N_1 & t_{12}/N_2 & \cdots & t_{1k}/N_k \\
    \vdots & \vdots & \ddots & \vdots \\
    t_{h1}/N_1 & t_{h2}/N_2 & \cdots & t_{hk}/N_k
    \end{pmatrix}.
    \]
    
    \State Solution subgroup: $A = (\mathrm{span}(T_h))^{\perp} \supseteq K^{\perp\perp} = K$;
    \State \Return $A$
  \end{algorithmic}
\end{algorithm}

In line~4 of Algorithm~\ref{algorithm1}, $0_G$ is the zero element of group $G$; $\mathcal{H}_{N_i}$ (for $1 \le i \le k$) is an $N_i$-dimensional Hilbert space with computational basis $\{\lvert x_i\rangle:x_i \in \{0, 1, \ldots, N_i-1\}\}$, and $\mathcal{H}_{G}$ is shorthand for $\mathop{\textstyle\bigotimes}\limits_{i=1}^k \mathcal{H}_{N_i}$, which is a $|G|$-dimensional Hilbert space. In lines 5 and 7 of Algorithm~\ref{algorithm1}, $\mathrm{I}_{G}:=\mathop{\textstyle\bigotimes}\limits_{i=1}^k \mathrm{I}_{N_i}$, where $\mathrm{I}_{N_i}$ is the identity operator on $\mathcal{H}_{N_i}$, and the quantum Fourier transform and its inverse are defined as:
\[
\mathrm{QFT}_{N_i} \lvert x_i \rangle = \frac{1}{\sqrt{N_i}} \sum_{y_i=0}^{N_i-1} e^{2\pi i x_i y_i / N_i} \lvert y_i \rangle,
\quad
\mathrm{QFT}_{N_i}^\dagger \lvert y_i \rangle = \frac{1}{\sqrt{N_i}} \sum_{x_i=0}^{N_i-1} e^{-2\pi i x_i y_i / N_i} \lvert x_i \rangle.
\]

We note that for general $N_i$, the $\mathrm{QFT}_{N_i}$ and $\mathrm{QFT}_{N_i}^\dagger$ cannot be implemented exactly without using high-dimensional qudits for the first registers~\cite{lomont2004hidden}. Hence, we adopt qudits, rather than qubits, to encode the elements of $G$.

In line~6, the oracle $U_f$ implements the mapping $U_f | g \rangle | b \rangle = | g \rangle | b + f(g) \rangle$ for all $g, b \in G$, where $+$ is the addition in Abelian group $G$.

In line 8 of Algorithm~\ref{algorithm1}, we obtain an element $\bm{t}_i=(t_{i1}, t_{i2},\dots, t_{ik})\in K^{\perp}$, and all elements $\bm{t}_i\in {K}^{\perp}$ are equally likely to be measured. After $h$ iterations, we can only guarantee that $\mathrm{span}(T_h)=\langle\bm{t}_1,\bm{t}_2,\cdots,\bm{t}_h\rangle\subseteq{K}^{\perp}$, rather than $\langle\bm{t}_1,\bm{t}_2,\cdots,\bm{t}_h\rangle={K}^{\perp}$. We then attempt to recover the hidden subgroup $K$ by solving the linear congruence system in line 11, with the aid of Smith normal form~\cite{lomont2004hidden, storjohann1996near}. The solution subgroup $A = \{\bm{x} \in G \mid \bm{W}\bm{x}^{\mathrm{T}}\equiv\bm{0}^{\mathrm{T}}\pmod{1}\}$ contains $K^{\perp\perp}$, where $K^{\perp\perp}:=(K^{\perp})^{\perp}$.

\begin{figure*}[h]
  \centering
  \includegraphics[scale=0.5]{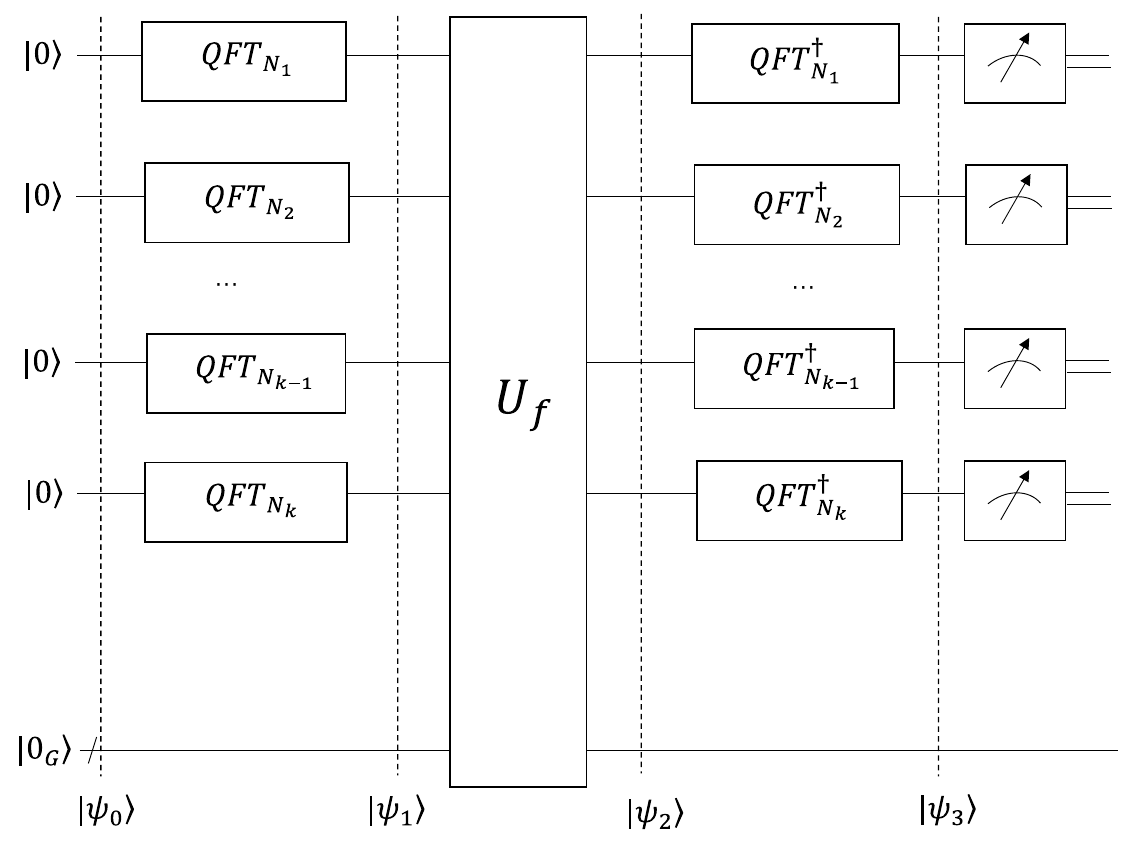}
  \caption{The circuit for Algorithm~\ref{algorithm1}.}
  \label{algorithm1_circuit}
\end{figure*}

The quantum circuit for AHSP is shown in Fig.~\ref{algorithm1_circuit}. Lines of quantum circuit in the figure represent  high-dimensional qudits, rather than qubits. In Fig.~\ref{algorithm1_circuit}, the first, second, $\ldots$ lines represent quantum systems of dimensions $N_1$, $N_2$, $\ldots$, respectively. The bottom line encodes the state $\lvert 0_G \rangle$, which belongs to the composite Hilbert space $\mathcal{H}_G=\mathop{\textstyle\bigotimes}\limits_{i=1}^k \mathcal{H}_{N_i}$.

Theorem~\ref{suanfachaxunfuzadu} guarantees $\langle\bm{t}_1,\bm{t}_2,\cdots,\bm{t}_h\rangle={K}^{\perp}$ with a probability exceed $1-\epsilon$. To precisely obtain $A=K$, we give an exact algorithm in Section~\ref{Sec4}, which requires only $\mathrm{len}(G)-\mathrm{len}(K)$ iterations.

\begin{thm}\label{suanfachaxunfuzadu}
In the quantum algorithm~\ref{algorithm1} for the finite AHSP, to ensure $\Pr(A = K) \ge 1-\epsilon$, it is sufficient to perform  
\[
h \ge \mathrm{rank}(G) + \left\lceil \log_2 \dfrac{2}{\epsilon} \right\rceil \quad \text{or} \quad h\ge \mathrm{len}(G) - \mathrm{len}(K) + \left\lceil \log_2 \dfrac{1}{\epsilon} \right\rceil.
\]
iterations.
\end{thm}
\begin{proof}

This follows directly from Proposition~\ref{zhihuolianchangdingli}. Note that ensuring \(\Pr(A = K) \ge 1 - \epsilon\) is equivalent to ensuring \(\Pr(\langle \mathbf{t}_1, \mathbf{t}_2, \dots, \mathbf{t}_h \rangle = K^\perp) = \varphi_h(K^\perp) \ge 1 - \epsilon\). By Proposition~\ref{zhihuolianchangdingli}, this probability holds if we choose:
\[
h \ge \mathrm{rank}(K^\perp) + \left\lceil \log_2 \frac{2}{\epsilon} \right\rceil \quad \text{or} \quad h \ge \mathrm{len}(K^\perp) + \left\lceil \log_2 \frac{1}{\epsilon} \right\rceil.
\]
Since $\mathrm{rank}(K^\perp)\le\mathrm{rank}(G)$ and $\mathrm{len}(K^\perp) = \mathrm{len}(G) - \mathrm{len}(K)$, 
we obtain the required bounds:

\[
h \ge \mathrm{rank}(G) + \left\lceil \log_2 \frac{2}{\epsilon} \right\rceil \quad \text{or} \quad h \ge \mathrm{len}(G) - \mathrm{len}(K) + \left\lceil \log_2 \frac{1}{\epsilon} \right\rceil.
\]
 
\end{proof}

\begin{remark}
Theorem~\ref{suanfachaxunfuzadu} shows the iteration count for success probability $1-\epsilon$ in the standard quantum algorithm for finite AHSP is $\mathrm{rank}(G) + \lceil \log_2 (2/\epsilon) \rceil$ or $\mathrm{len}(G) - \mathrm{len}(K) + \lceil \log_2 (1/\epsilon) \rceil$. The former offers an exponential improvement in $\epsilon$-dependence over the prior bound $\lfloor 4/\epsilon \rfloor \mathrm{rank}(G)$~\cite{koiran2007quantum}, while the latter improves upon $\lceil \log_2 |G| + \log_2 (1/\epsilon) + 2 \rceil$~\cite{lomont2004hidden}.
\end{remark}

\section{Exact quantum algorithm for finite AHSP}\label{Sec4}

\subsection{Application of quantum amplitude amplification to finite AHSP}

Quantum amplitude amplification provides a quadratic speedup over classical algorithms~\cite{BHMT02}. If we know the order $|K|$ of the hidden subgroup $K$ in AHSP, we can achieve exact algorithm for finite AHSP. Lemma~\ref{zhengjiaotonggoudingli} implies $|K^{\perp}| = |G|/|K|$, which is computed prior to execution. Consequently, the target proportion $\dfrac{|\mathrm{span}(T_{i-1})|}{|K^{\perp}|}$ is predetermined for
the $i$-th iteration in Algorithm~\ref{algorithm1}. By employing the method proposed by 
Brassard et al.~\cite{BHMT02}, we obtain an exact algorithm for AHSP, guaranteeing $A = K$. This approach requires at most $\mathrm{len}(G) - \mathrm{len}(K)$ iterations. We extend the method~\cite{BHMT02} to high-dimensional qudit system, rather than qubits.

In~\cite{imran2022exact}, a polynomial-time exact quantum algorithm for finite ASHP was proposed, but it is restricted to the additive group $G=\bigl(\mathbb{Z}_{m^k}\bigr)^n$ rather than general finite Abelian groups. They combine the Mihara--Sung variant of Brassard--Hoyer algorithm with a solution of a variant of key problem in the approach of Brassard and Hoyer. In this paper, we make an exact AHSP algorithm (EAHSP)  for any finite Abelian group $G$, and in a more concise way.

To make Algorithm~\ref{algorithm1} exact, we add a post-processing subroutine after line 7 of Algorithm~\ref{algorithm1}, to ensure the measured result $\bm{t}_i\notin \mathrm{span}(T_{i-1})$ in the $i$-th iteration. According to~\cite{kaye_introduction_2007}, in line 7 of Algorithm~\ref{algorithm1}, we have
\[
|\psi_3\rangle=\dfrac{1}{|K^{\perp}|}\sum_{\bm{t} \in K^{\perp}}\sum_{\bm{u}_r\in \Omega}e^{-2\pi i\langle\bm{t}, \bm{u}_r\rangle}|\bm{t}\rangle|f(\bm{u}_r)\rangle,
\]
where $\langle\cdot, \cdot\rangle$ is the bilinear form in Definition~\ref{shuangxianxingxingshi}.

The composite unitary operation from lines 5 to 7 in Algorithm~\ref{algorithm1} is given by
\[
\mathcal{A} = \left(\mathop{\textstyle\bigotimes}\limits_{i=1}^k \mathrm{QFT}_{N_i}^{\dagger} \otimes  \mathrm{I}_{G}\right)\cdot U_f \cdot \left(\mathop{\textstyle\bigotimes}\limits_{i=1}^k \mathrm{QFT}_{N_i} \otimes \mathrm{I}_{G}\right).
\]

In the $i$-th iteration, we define the phase rotation operators $\mathcal{R}_{0}(\phi_i)$ and $\mathcal{R}_{\mathcal{A}}(\varphi_i, T_{i-1})$ as follows:
\begin{equation}\label{eq:R_0}
\mathcal{R}_0(\phi_i)\Ket{a,b} = 
\begin{cases}
    \Ket{a,b}, & (a,b) \neq (0^k, 0_G) \in G \oplus G \\
    e^{\imath\phi_i}\Ket{a,b}, & (a,b) = (0^k, 0_G) \in G \oplus G 
\end{cases}
\end{equation}
where $\imath = \sqrt{-1}$, and
\begin{equation}\label{eq:R_A}
\mathcal{R}_{\mathcal{A}}(\varphi_i, T_{i-1})\Ket{a} = 
\begin{cases}
    e^{\imath\varphi_i}\Ket{a}, & a \notin \mathrm{span}(T_{i-1}) \\
    \Ket{a}, & a \in \mathrm{span}(T_{i-1}).
\end{cases}
\end{equation}

Using these operators, we construct the quantum amplitude amplification operator:
\begin{equation}\label{eq:Q}
\mathcal{Q}_i = \mathcal{A} \mathcal{R}_{0}(\phi_i) \mathcal{A}^{\dagger} \left(\mathcal{R}_{\mathcal{A}}(\varphi_i, T_{i-1}) \otimes \mathrm{I}_{G}\right).
\end{equation}

Let $Z_i = K^{\perp} \setminus \mathrm{span}(T_{i-1})$. We define the good subsets and bad subsets:
\begin{itemize}
    \item $\Ket{\Psi_{Z_i}}$: good subset spanned by $\{\Ket{a,b} \mid a \in Z_i, b \in G\}$ 
    \item $\Ket{\Psi_{Y_i}}$: bad subset spanned by $\{\Ket{a,b} \mid a \in \mathrm{span}(T_{i-1}), b \in G\}$
\end{itemize}
Note that $\Ket{\Psi_{Z_i}}$ and $\Ket{\Psi_{Y_i}}$ are not quantum states, as they are not normalized.

The state of $|\psi_3\rangle$ decomposes as:
\[
|\psi_3\rangle = \Ket{\Psi_{Z_i}} + \Ket{\Psi_{Y_i}}.
\]

The core idea for achieving an exact algorithm is to eliminate the amplitude of states in $\mathrm{span}(T_{i-1})$ in the $i$-th iteration. By appropriate selection of real parameters $\phi_i, \varphi_i \in \mathbb{R}$, a single application of $\mathcal{Q}_i$ to $|\psi_3\rangle$ suffices to achieve exact elimination. The quantum amplitude amplification procedure for measuring good states is described below. In Algorithm~\ref{algorithm2} and Theorem~\ref{jingqueziguochengdingli}, all indices $i$ refer to the $i$-th iteration.

\begin{algorithm}[t]
  \caption{Quantum amplitude amplification for measuring good states (QAA)}\label{algorithm2}
  \begin{algorithmic}[1]
    \Statex \textbf{Input:} Registers $|\psi_3\rangle$, operator $\mathcal{A}$, set $T_{i-1}$
    \Statex \textbf{Output:} $\bm{t}_i\in K^{\perp}\setminus \mathrm{span}(T_{i-1})$
    
    \State $b_i\gets 1-\dfrac{|\mathrm{span}(T_{i-1})| \cdot |K|}{|G|}$;
    \State $\phi_i\gets \arccos(1-\dfrac{1}{2b_i})$ or $2\pi-\arccos(1-\dfrac{1}{2b_i})$;
    \State $\varphi_i\gets \arccos(1-\dfrac{1}{2b_i})$ or $2\pi-\arccos(1-\dfrac{1}{2b_i})$;
    \State Apply $\mathcal{Q}_i$ once to $|\psi_3\rangle$ to get $|\psi_4\rangle=\mathcal{Q}_i|\psi_3\rangle$, where $\mathcal{Q}_i=\mathcal{A}\mathcal{R}_{0}(\phi_i)\mathcal{A}^{\dagger}\left(\mathcal{R}_{\mathcal{A}}(\varphi_i, T_{i-1})\otimes \mathrm{I}_{G}\right)$;
    \State Measure the first register to obtain $\bm{t}_i=(t_{i1}, t_{i2},\dots, t_{ik})\in K^{\perp}\setminus \mathrm{span}(T_{i-1})$;
    \State \Return $\bm{t}_i$
  \end{algorithmic}
\end{algorithm}

Theorem~\ref{jingqueziguochengdingli} guarantees the correctness of Algorithm~\ref{algorithm2}.

\begin{thm}\label{jingqueziguochengdingli} 
Let $\varphi_i=\phi_i=\arccos(1-\dfrac{1}{2b_i})$ or $\varphi_i=\phi_i=2\pi-\arccos(1-\dfrac{1}{2b_i})$, where $b_i = 1 - \dfrac{|\mathrm{span}(T_{i-1})| \cdot |K|}{|G|}$. Then
\[
\mathcal{Q}_i|\psi_3\rangle =\Ket{\Psi_{Z_i}}.
\]

Hence, line 5 of Algorithm~\ref{algorithm2} deterministically yields $\bm{t}_i \in K^\perp \setminus \mathrm{span}(T_{i-1})$.

\end{thm}

\begin{proof}
From Eq.~(\ref{eq:R_0}), we can write $\mathcal{R}_{0}(\phi_i)$ as follows,
\begin{equation*}
    \mathcal{R}_{0}(\phi_i) =\mathrm{I}_{G}\otimes \mathrm{I}_{G}- \left(1 - e^{\imath\phi_i}\right)\Ket{0^{k}, 0_G}\Bra{0^{k}, 0_G}.
\end{equation*}

From the definitions of $\mathcal{R}_{\mathcal{A}}(\varphi_i, T_{i-1})$, $\Ket{\Psi_{Z_i}}$, and $\Ket{\Psi_{Y_i}}$, we have
\begin{align*}
    \left(\mathcal{R}_{\mathcal{A}}(\varphi_i, T_{i-1})\otimes \mathrm{I}_{G}\right)\Ket{\Psi_{Z_i}} &= e^{\imath\varphi_i}\Ket{\Psi_{Z_i}}, \\
    \left(\mathcal{R}_{\mathcal{A}}(\varphi_i, T_{i-1})\otimes \mathrm{I}_{G}\right)\Ket{\Psi_{Y_i}} &= \Ket{\Psi_{Y_i}}.
\end{align*}

Let $\mathcal{U}(\mathcal{A}, \phi_i) = \mathcal{A}\mathcal{R}_{0}(\phi_i)\mathcal{A}^{\dagger}$, then based on Eq.~(\ref{eq:Q}), $\mathcal{Q}_i$ can be written as
\begin{equation*}
    \mathcal{Q}_i = \mathcal{U}(\mathcal{A}, \phi_i)\left(\mathcal{R}_{\mathcal{A}}(\varphi_i, T_{i-1})\otimes \mathrm{I}_{G}\right).
\end{equation*}

For $\mathcal{U}(\mathcal{A}, \phi_i)$, we have
\begin{align*}
\mathcal{U}(\mathcal{A}, \phi_i) =&\,\mathcal{A}\mathcal{R}_{0}(\phi_i)\mathcal{A}^{\dagger} \\
               =&\,\mathcal{A}\left(\mathrm{I}_{G}\otimes \mathrm{I}_{G} - \left(1 - e^{\imath\phi_i}\right) \Ket{0^{k}, 0_G} \Bra{0^{k}, 0_G}\right) \mathcal{A}^{\dagger} \\
               =&\,-\left(1 - e^{\imath\phi_i}\right)\left(\mathcal{A}\Ket{0^{k}, 0_G}\Bra{0^{k}, 0_G}\mathcal{A}^{\dagger}\right)+\mathrm{I}_{G}\otimes \mathrm{I}_{G} \\
              =&\,-\left(1 - e^{\imath\phi_i}\right)|\psi_3\rangle\langle\psi_3|+ \mathrm{I}_{G}\otimes \mathrm{I}_{G} \\
               =&\,-\left(1 - e^{\imath\phi_i}\right)(\Ket{\Psi_{Z_i}}+\Ket{\Psi_{Y_i}})(\Bra{\Psi_{Z_i}}+\Bra{\Psi_{Y_i}})+ \mathrm{I}_{G}\otimes \mathrm{I}_{G}.
\end{align*}

Prior to applying amplitude amplification, we require the strict condition $|\mathrm{span}(T_{i-1})| < |K^{\perp}|$ (rather than $\leq$). Given that $|K^{\perp} | =|G|/|K|$, and using the definitions of $\ket{\Psi_{Z_i}}$ and $\ket{\Psi_{Y_i}}$, we have
\begin{align*}
    \Braket{\Psi_{Z_i}|\Psi_{Z_i}} &=1-\frac{|\mathrm{span}(T_{i-1})|}{|K^{\perp}|}= 1 - \frac{|\mathrm{span}(T_{i-1})|\cdot|K|}{|G|}, \\
    \Braket{\Psi_{Y_i}|\Psi_{Y_i}} &=\frac{|\mathrm{span}(T_{i-1})|}{|K^{\perp}|}= \frac{|\mathrm{span}(T_{i-1})|\cdot|K|}{|G|}, \\
    \Braket{\Psi_{Z_i}|\Psi_{Y_i}} &= 0.
\end{align*}

Thus, we obtain the following equations.
\begin{align*}
\mathcal{Q}_i\Ket{\Psi_{Z_i}} &= \mathcal{U}(\mathcal{A}, \phi_i)\left(\mathcal{R}_{\mathcal{A}}(\varphi_i, T_{i-1})\otimes \mathrm{I}_{G}\right)\Ket{\Psi_{Z_i}} \\
&= e^{\imath\varphi_i}\mathcal{U}(\mathcal{A}, \phi_i)\Ket{\Psi_{Z_i}} \\
&= e^{\imath\varphi_i}\bigg[-\left(1 - e^{\imath\phi_i}\right)\big(\Ket{\Psi_{Z_i}}+\Ket{\Psi_{Y_i}}\big)\big(\Bra{\Psi_{Z_i}}+\Bra{\Psi_{Y_i}}\big) + \mathrm{I}_{G}\otimes\mathrm{I}_{G}\bigg]\Ket{\Psi_{Z_i}} \\
&= -e^{\imath\varphi_i}\left(1 - e^{\imath\phi_i}\right)\big(\Ket{\Psi_{Z_i}}+\Ket{\Psi_{Y_i}}\big)\Braket{\Psi_{Z_i}|\Psi_{Z_i}} + e^{\imath\varphi_i}\Ket{\Psi_{Z_i}} \\
&= -e^{\imath\varphi_i}\left(1 - e^{\imath\phi_i}\right)\left(1 - \frac{|\mathrm{span}(T_{i-1})|\cdot|K|}{|G|}\right)\big(\Ket{\Psi_{Z_i}}+\Ket{\Psi_{Y_i}}\big) + e^{\imath\varphi_i}\Ket{\Psi_{Z_i}} \\
&= -e^{\imath\varphi_i}\left[\left(1 - e^{\imath\phi_i}\right)\left(1 - \frac{|\mathrm{span}(T_{i-1})|\cdot|K|}{|G|}\right)-1\right]\Ket{\Psi_{Z_i}} \\
&\quad -e^{\imath\varphi_i}\left(1 - e^{\imath\phi_i}\right)\left(1 - \frac{|\mathrm{span}(T_{i-1})|\cdot|K|}{|G|}\right)\Ket{\Psi_{Y_i}}.					  
\end{align*}

\begin{align*}
    \mathcal{Q}_i\Ket{\Psi_{Y_i}} =&\, \mathcal{U}(\mathcal{A}, \phi_i)\left(\mathcal{R}_{\mathcal{A}}(\varphi_i,T_{i-1})\otimes \mathrm{I}_{G}\right)\Ket{\Psi_{Y_i}} \\
                              =&\, \mathcal{U}(\mathcal{A}, \phi_i)\Ket{\Psi_{Y_i}} \\
                              =&\, \bigg[-\left(1 - e^{\imath\phi_i}\right)\big(\Ket{\Psi_{Z_i}}+\Ket{\Psi_{Y_i}}\big)\big(\Bra{\Psi_{Z_i}}+\Bra{\Psi_{Y_i}}\big)+\mathrm{I}_{G}\otimes \mathrm{I}_{G}\bigg]\Ket{\Psi_{Y_i}} \\
                              =&\, -\left(1 - e^{\imath\phi_i}\right)\big(\Ket{\Psi_{Z_i}}+\Ket{\Psi_{Y_i}}\big)\Braket{\Psi_{Y_i}|\Psi_{Y_i}} + \Ket{\Psi_{Y_i}} \\
                              =&\,-\left(1 - e^{\imath\phi_i}\right)\frac{|\mathrm{span}(T_{i-1})|\cdot|K|}{|G|}\big(\Ket{\Psi_{Z_i}}+\Ket{\Psi_{Y_i}}\big) + \Ket{\Psi_{Y_i}} \\
                              =&\,-\left(1 - e^{\imath\phi_i}\right)\frac{|\mathrm{span}(T_{i-1})|\cdot|K|}{|G|}\Ket{\Psi_{Z_i}}+ \left[\left(1 - e^{\imath\phi_i}\right)\left(1 - \frac{|\mathrm{span}(T_{i-1})|\cdot|K|}{|G|}\right) + e^{\imath\phi_i}\right]\Ket{\Psi_{Y_i}}.						  
\end{align*}

To eliminate bad state amplitudes in $\Ket{\Psi_{Y_i}}$ after applying $Q_i$ to $\Ket{\Psi_{Z_i}}+\Ket{\Psi_{Y_i}}$, we must guarantee
\begin{equation}\label{zhenfuxiaoqu}
    e^{\imath\varphi_i}\left(1 - e^{\imath\phi_i}\right)\left(1 - \frac{|\mathrm{span}(T_{i-1})|\cdot|K|}{|G|}\right) =\left(1 - e^{\imath\phi_i}\right)\left(1 - \frac{|\mathrm{span}(T_{i-1})|\cdot|K|}{|G|}\right) + e^{\imath\phi_i}.
\end{equation}

Denote
\begin{equation*}
    b_i = 1 - \dfrac{|\mathrm{span}(T_{i-1})|\cdot|K|}{|G|}.
\end{equation*}

As $|\mathrm{span}(T_{i-1})| < |K^{\perp}|$ (rather than $\leq$), it follows directly that $\dfrac{1}{2} \le b_i \le 1$. Dividing both sides of Eq.~(\ref{zhenfuxiaoqu}) by $e^{\imath\varphi_i}\left(1 - e^{\imath\phi_i}\right) \neq 0$, we obtain
\begin{align}
    b_i &= \dfrac{e^{\imath\phi_i}}{(1-e^{\imath\phi_i})(e^{\imath\varphi_i}-1)} \nonumber\\  
      &= \dfrac{e^{\imath\phi_i}(1-e^{-\imath\phi_i})(e^{-\imath\varphi_i}-1)}{|1-e^{\imath\phi_i}|^2|e^{\imath\varphi_i}-1|^2} \nonumber\\
      &= \dfrac{(e^{\imath\phi_i}-1)(e^{-\imath\varphi_i}-1)}{|1-e^{\imath\phi_i}|^2|e^{\imath\varphi_i}-1|^2} \nonumber\\
      &= \dfrac{e^{\imath(\phi_i-\varphi_i)} - e^{\imath\phi_i} - e^{-\imath\varphi_i} + 1}{|1-e^{\imath\phi_i}|^2|e^{\imath\varphi_i}-1|^2}.\label{eq:varphi}				  
\end{align}

Since $b_i$ is real, the imaginary part of Eq.~(\ref{eq:varphi})'s numerator must vanish. Using $e^{\imath x} = \cos x + \imath \sin x$ and extracting the imaginary part gives:
\begin{equation*}
\sin(\phi_i-\varphi_i) - \sin\phi_i + \sin\varphi_i = 0
\end{equation*}
which simplifies to:
\begin{equation}
\sin(\phi_i - \varphi_i) = \sin\phi_i - \sin\varphi_i.\label{eq:im_eq}
\end{equation}

Applying the sum-to-product identity to the right-hand side of Eq.~(\ref{eq:im_eq}):
\begin{equation}
\sin\phi_i - \sin\varphi_i = 2\cos\left(\frac{\phi_i+\varphi_i}{2}\right)\sin\left(\frac{\phi_i-\varphi_i}{2}\right).\label{eq:sum_prod}
\end{equation}

The left-hand side of Eq.~(\ref{eq:im_eq}) can be rewritten using the double-angle formula:
\begin{equation}
\sin(\phi_i - \varphi_i) = 2\sin\left(\frac{\phi_i-\varphi_i}{2}\right)\cos\left(\frac{\phi_i-\varphi_i}{2}\right).\label{eq:double_angle}
\end{equation}

Equating the expressions from \eqref{eq:sum_prod} and \eqref{eq:double_angle} via Eq.~(\ref{eq:im_eq}):
\begin{equation*}
2\sin\left(\frac{\phi_i-\varphi_i}{2}\right)\cos\left(\frac{\phi_i-\varphi_i}{2}\right) = 2\cos\left(\frac{\phi_i+\varphi_i}{2}\right)\sin\left(\frac{\phi_i-\varphi_i}{2}\right),
\end{equation*}
which simplifies to:
\begin{equation}
\left[\cos\left(\frac{\phi_i-\varphi_i}{2}\right) - \cos\left(\frac{\phi_i+\varphi_i}{2}\right)\right]\sin\left(\frac{\phi_i-\varphi_i}{2}\right) = 0.\label{eq:simplified}
\end{equation}

Applying the product-to-sum identity $\cos A - \cos B = -2\sin\left(\frac{A+B}{2}\right)\sin\left(\frac{A-B}{2}\right)$, Eq.~(\ref{eq:simplified}) becomes:
\begin{equation}
-2\sin\left(\frac{\phi_i}{2}\right)\sin\left(-\frac{\varphi_i}{2}\right)\sin\left(\frac{\phi_i-\varphi_i}{2}\right) = 0.\label{sanjiaohanshu}
\end{equation}

Since $\phi_i\in[0,2\pi)$ and $\varphi_i\in[0,2\pi)$, we have $\dfrac{\phi_i-\varphi_i}{2}\in (-\pi,\pi)$. 
From Eq.~(\ref{sanjiaohanshu}), we deduce three possibilities: $\phi_i=0$, $\varphi_i=0$, or $\phi_i=\varphi_i$.
However, this is only a necessary condition derived from the vanishing imaginary part. Substituting into Eq.~(\ref{zhenfuxiaoqu}) confirms that only $\phi_i=\varphi_i$ is valid. With $\phi_i=\varphi_i$, Eq.~(\ref{eq:varphi}) simplifies to:
\begin{equation*}
b_i=-\frac{e^{\imath\phi_i}}{(1-e^{\imath\phi_i})^2}.
\end{equation*}

Solving this equation for $e^{\imath\phi_i}$ yields:

\begin{equation}
e^{\imath\phi_i}=\dfrac{2-\dfrac{1}{b_i}+\sqrt{\dfrac{1}{{b_i}^2}-\dfrac{4}{b_i}}}{2},\label{zhishuhanshu}
\end{equation}
which is equivalent to 
\begin{align}
\phi_i&=-\imath\mathrm{Ln}\left(1-\frac{1}{2b_i}+\sqrt{\frac{1}{4{b_i}^2}-\frac{1}{b_i}}\right)\label{duishuhanshu}\\
&=\mathrm{Arccos}\left(1-\frac{1}{2b_i}\right)
\quad\text{\Big(since $\mathrm{Arccos}(z) = -\imath\mathrm{Ln}(z + \sqrt{z^2-1}),\ \forall z\in\mathbb{C}$\Big)}\label{dasanjiaohanshu}\\
&=\pm\arccos\left(1 - \frac{1}{2b_i}\right)+2k\pi\quad(k\in \mathbb{Z})\label{xiaosanjiaohanshu}.
\end{align}

Equations~(\ref{zhishuhanshu})-(\ref{xiaosanjiaohanshu}) are multi-valued complex functions. The square roots in Eqs.~(\ref{zhishuhanshu}) and (\ref{duishuhanshu}) are double-valued (e.g., $\sqrt{4}=\pm 2$), while $\mathrm{Ln}(z)$ and $\mathrm{Arccos}(z)$ are multi-valued. For all $z\in \mathbb{C}$, $\arccos(z)$ denotes the principal value branch of $\mathrm{Arccos}(z)$, satisfying
\[
\arccos(z) \in \left\{ w \in \mathbb{C} \mid \operatorname{Re}(w) \in [0, \pi],\  \operatorname{Im}(w) \in \mathbb{R} \right\}.
\]

Given $\dfrac{1}{2} \leq b_i \leq 1$, the quantity $1 - \dfrac{1}{2b_i}$ lies in $[0, \dfrac{1}{2}]$. Consequently, $\arccos\left(1 - \dfrac{1}{2b_i}\right)$ is real-valued. Combining this with the phase constraint $\phi_i \in [0, 2\pi)$ yields two solutions:

\[
\phi_i = \arccos\left(1 - \dfrac{1}{2b_i}\right) \quad \text{or} \quad \phi_i = 2\pi - \arccos\left(1 - \dfrac{1}{2b_i}\right).
\]

The corresponding phase angles satisfy
\[
\phi_i \in \left[\dfrac{\pi}{3}, \dfrac{\pi}{2}\right] \cup \left[\dfrac{3\pi}{2}, \dfrac{5\pi}{3}\right].
\]

The final analytical expressions are therefore:
\begin{align*}
\phi_i = \varphi_i &= \arccos\left(1 - \frac{1}{2b_i}\right) \\
&= \arccos\left(1 - \frac{1}{2\left(1 - \dfrac{|\mathrm{span}(T_{i-1})| \cdot |K|}{|G|}\right)}\right), \\
\intertext{or}
\phi_i = \varphi_i &= 2\pi - \arccos\left(1 - \frac{1}{2b_i}\right) \\
&= 2\pi - \arccos\left(1 - \frac{1}{2\left(1 - \dfrac{|\mathrm{span}(T_{i-1})| \cdot |K|}{|G|}\right)}\right).
\end{align*}

\end{proof}

\subsection{Constructing exact quantum algorithm for finite AHSP}

\begin{figure}[H]
  \centering
  \includegraphics[scale=0.46]{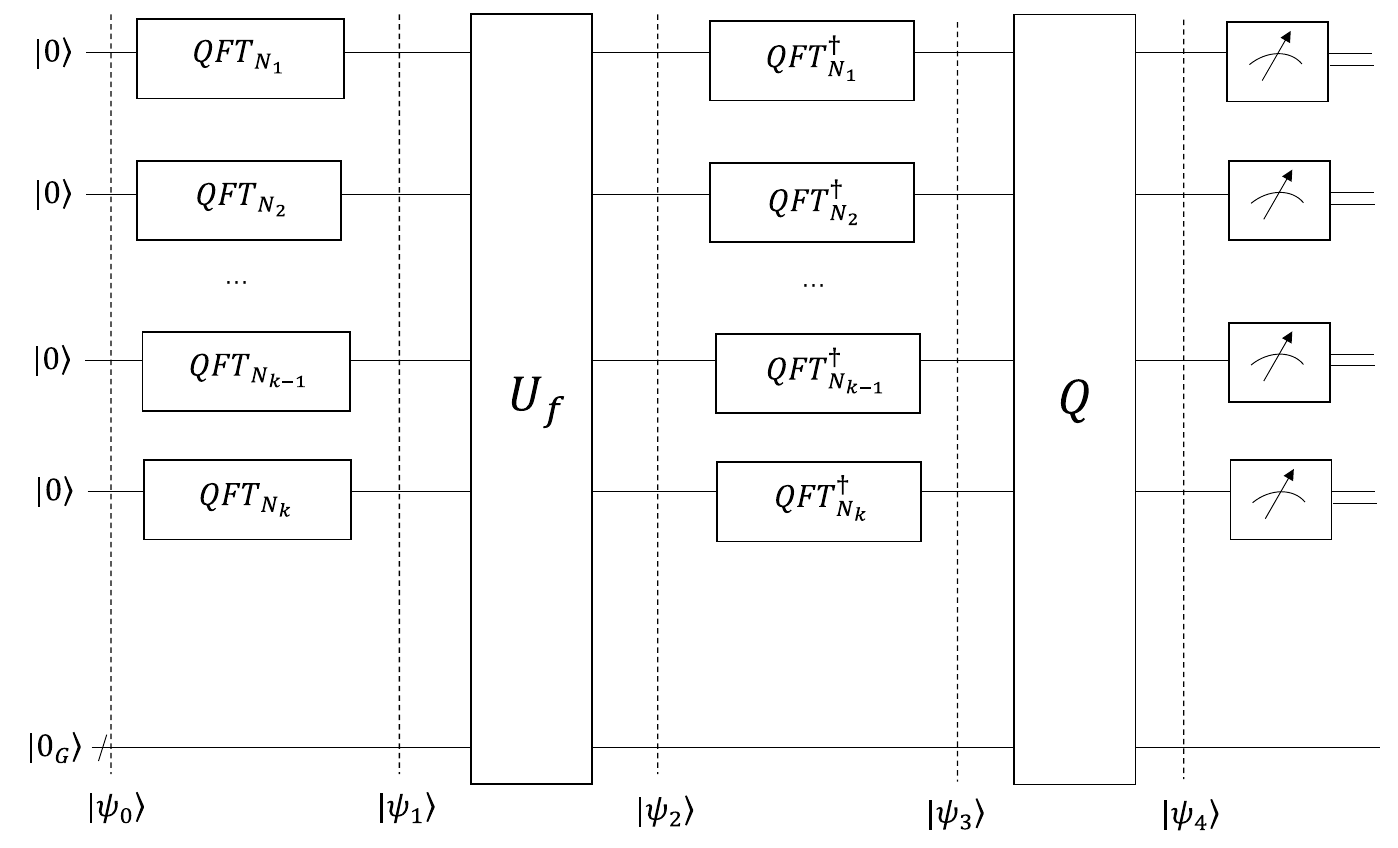}
  \caption{The circuit for  EAHSP (Algorithm~\ref{algorithm3}).}
  \label{algorithm3_circuit}
\end{figure}

We introduce Algorithm~\ref{algorithm3} (EAHSP), an exact algorithm for the AHSP based on the amplitude amplification procedure of Algorithm~\ref{algorithm2}. In lines 4 and 5 of Algorithm~\ref{algorithm3},  $0_G$ is the zero element of group $G$, $\mathrm{I}_{G}:=\mathop{\textstyle\bigotimes}\limits_{i=1}^k \mathrm{I}_{N_i}$, $\mathcal{H}_G:=\mathop{\textstyle\bigotimes}\limits_{i=1}^k \mathcal{H}_{N_i}$.

The quantum circuit for EAHSP is shown in Fig.~\ref{algorithm3_circuit}. In the bottom line, $\lvert 0_G \rangle\in \mathcal{H}_G$.

\begin{algorithm}[H]
  \caption{Exact quantum algorithm for finite AHSP (EAHSP)}\label{algorithm3}
  \begin{algorithmic}[1]
    \Statex \textbf{Input:} Integer $N_1,N_2,\cdots,N_k$, oracle $U_f$
    \Statex \textbf{Output:} Subgroup $A = K$ (exact solution)
    
    \State $T_0 \gets \emptyset$; 
    \State $h\gets\mathrm{len}(G)-\mathrm{len}(K)$; \Comment{By Theorem~\ref{jingquechaxuncishu}}
    \For{$i = 1$ to $h$}
 \State $|\psi_0\rangle=|0\rangle|0\rangle \cdots|0\rangle |0_G\rangle \in \mathcal{H}_{N_1} \otimes \mathcal{H}_{N_2} \otimes \cdots \otimes \mathcal{H}_{N_k} \otimes \mathcal{H}_{G}$;
      \State Apply $\mathcal{A}$ to obtain $|\psi_3\rangle=\mathcal{A}|\psi_0\rangle$, where $\mathcal{A} = (\mathop{\textstyle\bigotimes}\limits_{i=1}^k \mathrm{QFT}_{N_i}^{\dagger} \otimes  \mathrm{I}_{G})\cdot U_f \cdot (\mathop{\textstyle\bigotimes}\limits_{i=1}^k \mathrm{QFT}_{N_i} \otimes \mathrm{I}_{G})$;
      \State $\bm{t}_i \gets$ \Call{QAA}{$|\psi_3\rangle$, $\mathcal{A}$, $T_{i-1}$}; \Comment{QAA by Algorithm~\ref{algorithm2}}
      \State Update $T_i\gets T_{i-1}\cup \{\bm{t}_i\}$; 
    \EndFor
    
    \State Solve linear congruence system:
    \[
    \bm{W}\bm{x}^{\mathrm{T}} \equiv \bm{0}^{\mathrm{T}} \pmod{1} \quad \text{where } \bm{W}=\begin{pmatrix} 
    t_{11}/N_1 & t_{12}/N_2 & \cdots & t_{1k}/N_k \\
    \vdots & \vdots & \ddots & \vdots \\
    t_{h1}/N_1 & t_{h2}/N_2 & \cdots & t_{hk}/N_k
    \end{pmatrix}.
    \]
    
    \State Solution subgroup: $A = (\mathrm{span}(T_h))^{\perp}= K^{\perp\perp} = K$;
    \State \Return $A$
  \end{algorithmic}
\end{algorithm}

\begin{thm}\label{jingquechaxuncishu}
\begin{enumerate}[label=\textup{(\roman*)}, itemsep=0pt,align=left,labelwidth=1.5em]
    \item In Algorithm~\ref{algorithm3}, after performing $\mathrm{len}(G)-\mathrm{len}(K)$ iterations, we can exactly obtain $A=K$;
    \item The quantum queries to oracle $U_f$ is $3(\mathrm{len}(G)-\mathrm{len}(K))$, thus the asymptotic complexity of Algorithm~\ref{algorithm3} is $\mathcal{O}(\mathrm{len}(G)-\mathrm{len}(K))$.
\end{enumerate}

\end{thm}

\begin{proof}

(i) In line 5 of Algorithm~\ref{algorithm3}, we get the state $|\psi_3\rangle=\mathcal{A}|\psi_0\rangle$. In 
line 6 (QAA subroutine), applying $\mathcal{Q}_i$ on $|\psi_3\rangle$ yields $|\psi_4\rangle=\mathcal{Q}_i|\psi_3\rangle$, where $\mathcal{Q}_i=\mathcal{A}\mathcal{R}_{0}(\phi_i)\mathcal{A}^{\dagger}\left(\mathcal{R}_{\mathcal{A}}(\varphi_i, T_{i-1})\otimes \mathrm{I}_{G}\right)$, $\mathcal{A} = (\mathop{\textstyle\bigotimes}\limits_{i=1}^k \mathrm{QFT}_{N_i}^{\dagger} \otimes  \mathrm{I}_{G})\cdot U_f \cdot (\mathop{\textstyle\bigotimes}\limits_{i=1}^k \mathrm{QFT}_{N_i} \otimes \mathrm{I}_{G})$, $\phi_i=\varphi_i=\arccos(1-\dfrac{1}{2b_i})$ or $\phi_i=\varphi_i=2\pi-\arccos(1-\dfrac{1}{2b_i})$, $b_i=1 - \dfrac{|\mathrm{span}(T_{i-1})|\cdot|K|}{|G|}$.

Theorem~\ref{jingqueziguochengdingli} implies that $|\psi_4\rangle = \left| \Psi_{Z_i} \right\rangle$. Therefore, measuring the first register of \(|\psi_4\rangle\) in line 6 of Algorithm~\ref{algorithm3} yields an element $\bm{t}_i = (t_{i1}, t_{i2}, \dots, t_{ik}) \in K^{\perp} \setminus \mathrm{span}(T_{i-1})$. This ensures strict growth of the subgroup chain: $\mathrm{span}(T_{i-1}) \subsetneqq \mathrm{span}(T_i)$. 

Thus after completing all $h = \mathrm{len}(G) - \mathrm{len}(K)$ iterations in Algorithm~\ref{algorithm3}, we have $\mathrm{len}(\mathrm{span}(T_h)) \ge \mathrm{len}(K^\perp) = \mathrm{len}(G) - \mathrm{len}(K)$, which implies $\mathrm{span}(T_h) = K^{\perp}$. Finally, in line 10, solving the system of congruences gives $A=(\mathrm{span}(T_h))^{\perp}=K$.

(ii) In the $i$-th iteration of Algorithm~\ref{algorithm3}, the oracle $U_f$ is queried three times: \textbf{once} by the operator $\mathcal{A}$ in line 5, and \textbf{twice} during the QAA subroutine in line 6 (corresponding to applications of $\mathcal{A}$ and $\mathcal{A}^{\dagger}$). Overall, the total quantum queries to oracle $U_f$ is $3(\mathrm{len}(G)-\mathrm{len}(K))$. Thus the asymptotic complexity of Algorithm~\ref{algorithm3} is $\mathcal{O}(\mathrm{len}(G)-\mathrm{len}(K))$.
\end{proof}

\begin{remark}
In the best case, the exact algorithm terminates within $\mathrm{rank}(K^\perp)$ iterations. However, Proposition~\ref{zhiguanxi} shows that $\mathrm{rank}(K^{\perp}) \ge \mathrm{rank}(G) - \mathrm{rank}(K)$, meaning the iteration count may exceed $\mathrm{rank}(G) - \mathrm{rank}(K)$. 

Since $\mathrm{rank}(G)$ and $\mathrm{rank}(K)$ alone cannot bound the iteration complexity, we use chain length in Algorithm~\ref{algorithm3}. In the worst case, $\mathrm{len}(G) - \mathrm{len}(K)$ iterations are necessary and sufficient.
\end{remark}

\begin{remark}
Algorithm~\ref{algorithm3} is more concise than the previous exact algorithm~\cite{imran2022exact} for finite AHSP. Exact algorithm in~\cite{imran2022exact} is only applicable to groups $G=\bigl(\mathbb{Z}_{m^k}\bigr)^n$ and requires $\mathcal{O}\big(nk \cdot \log^{2} m\big)$ quantum queries, equivalent to $\mathcal{O}\big(\mathrm{len}(G) \log^{2} m\big)$. In contrast, Algorithm~\ref{algorithm3} applies to any finite Abelian group $G$ and achieves a significantly lower complexity of $3(\mathrm{len}(G) - \mathrm{len}(K))$. This complexity is independent of $m$ and is further reduced by incorporating the term $\mathrm{len}(K)$.
\end{remark}

\section{Distributed quantum algorithms for finite AHSP}\label{Sec5}

In this section, we present the distributed quantum algorithms for finite AHSP. Before that, we give two theorems serving as the foundation of our distributed quantum algorithm.

\subsection{Subgroup decomposition in direct products and sums}

We present Theorem~\ref{Subgroup Decomposition via Direct Products}, applicable to distributed systems over non-Abelian groups, and Theorem~\ref{External Direct Sum Decomposition of Subgroups} as its corollary. Part (i) of Theorem~\ref{External Direct Sum Decomposition of Subgroups} establishes the theoretical foundation for classical distributed algorithms, while the combination of parts (i) and (ii) ensures the correctness of quantum distributed algorithms for finite AHSP in this work.

Both Theorems~\ref{Subgroup Decomposition via Direct Products} and~\ref{External Direct Sum Decomposition of Subgroups} are formulated using external direct products. While internal and external direct products are isomorphic, the external formulation better suits our distributed structures, as it reduces qudit number—critical for quantum implementations.

\begin{thm}[\textbf{Subgroup Decomposition via Direct Products}]\label{Subgroup Decomposition via Direct Products}
Let $G$ be a finite group with external direct product $G = G_1 \times G_2 \times \cdots \times G_m$ 
where $\gcd(|G_i|, |G_j|) = 1$ for all $1 \leq i < j \leq m$.

Then for any subgroup $K \leq G$, there exist \textbf{unique} subgroups $K_i \leq G_i(1 \leq i \leq m)$ such that
\[
K = K_1 \times K_2 \times \cdots \times K_m
\]
as an external direct product.
\end{thm}

\begin{proof}

The proof establishes both existence and uniqueness of the subgroups $K_i$, primarily employing \textbf{Chinese Remainder Theorem}.

\textbf{Step 1: Projection maps and subgroup structure}

Let $G = \displaystyle\prod_{i=1}^m G_i$ and define the projection maps:
\[
\pi_i: G \to G_i, \quad (g_1, \dots, g_m) \mapsto g_i \quad (1 \leq i \leq m).
\]
For any subgroup $K \leq G$, set $K_i = \pi_i(K)$. To verify $K_i \leq G_i$:
\begin{itemize}
    \item \textit{Identity}: $(e_1, \dots, e_m) \in K \Rightarrow e_i \in K_i$, where $e_i$ is the identity in $G_i$.
    \item \textit{Closure}: For $a, b \in K_i$, take $x, y \in K$ with $\pi_i(x) = a$, $\pi_i(y) = b$. Then
    \[
    ab^{-1} = \pi_i(x)(\pi_i(y))^{-1} = \pi_i(xy^{-1}) \in K_i
    \]
    since $xy^{-1} \in K$.
\end{itemize}
Thus $K_i \leq G_i$.

\textbf{Step 2: $K \subseteq \displaystyle\prod_{i=1}^m K_i$}

For any $k = (k_1, \dots, k_m) \in K$, we have $k_i = \pi_i(k) \in \pi_i(K) = K_i$ for each $i$. 
Thus $k\in\displaystyle\prod_{i=1}^m K_i$, establishing the inclusion.

\textbf{Step 3: $\displaystyle\prod_{i=1}^m K_i \subseteq K$}

Let $n_i = |G_i|$. By \textbf{the Chinese Remainder Theorem} (since $n_i$ are pairwise coprime), there exist integers $c_i$ such that:
\[
c_i \equiv 
\begin{cases} 
1 \pmod{n_i}, \\ 
0 \pmod{n_j}, & j \neq i.
\end{cases}
\]

For each $k_i \in K_i$, choose $x^{(i)} = (a_1^{(i)}, \dots, k_i, \dots, a_m^{(i)}) \in K$ with $\pi_i(x^{(i)}) = k_i$.

Consider $(x^{(i)})^{c_i} \in K$. By the congruence conditions:
\begin{itemize}
    \item For $j \neq i$: $n_j \mid c_i \Rightarrow (a_j^{(i)})^{c_i} = e_j$ (identity of $G_j$).
    \item For component $i$: $c_i \equiv 1 \pmod{n_i} \Rightarrow (k_i)^{c_i} = k_i$.
\end{itemize}
Thus $(x^{(i)})^{c_i} = (e_1, \dots, k_i, \dots, e_m) \in K$.

For any $(k_1, \dots, k_m) \in \displaystyle\prod_{i=1}^m K_i$, we have $(k_1, \dots, k_m)= \displaystyle\prod_{i=1}^m (e_1, \dots, e_{i-1}, k_i, e_{i+1},\dots, e_m)=\displaystyle\prod_{i=1}^m (x^{(i)})^{c_i} \in K$, establishing the inclusion.

Combining with Step 2, we conclude $K  = K_1 \times K_2 \times \cdots \times K_m$.

\textbf{Step 4: Uniqueness of the decomposition}

The subgroups $K_i$ are uniquely determined as $K_i = \pi_i(K)$. Indeed, if $K = L_1 \times \cdots \times L_m$ were another decomposition with $L_i \leq G_i$, then applying $\pi_i$ yields:
\[
\pi_i(K) = \pi_i(L_1 \times \cdots \times L_m) = L_i,
\]
so $L_i = K_i$ for all $i$.
\end{proof}

\begin{remark}
A typical example of group $G$ in Theorem~\ref{Subgroup Decomposition via Direct Products} is \textbf{finite nilpotent} groups, with the factors $G_i$ being their Sylow $p_i$-subgroups.

Theorem~\ref{Subgroup Decomposition via Direct Products}, along with part (i) of Algorithm~\ref{External Direct Sum Decomposition of Subgroups}, does not hold for the group $G = \mathbb{Z}_2^n$ in \textbf{Simon's problem} due to the lack of pairwise coprime direct factors.

\end{remark}

\begin{thm}[\textbf{Abelian Subgroup Decomposition via Direct Sum}]\label{External Direct Sum Decomposition of Subgroups}
Let $G$ be a finite Abelian group with external direct sum decomposition
$G = G_1 \oplus G_2 \oplus \cdots \oplus G_m$,
where:
\begin{itemize}
    \item Each $\widetilde{G}_i = \{0_1\} \times \cdots \times G_i \times \cdots \times \{0_m\}$ is the Sylow $p_i$-subgroup of $G$;
    \item The primes $p_1, \ldots, p_m$ are distinct;
    \item $0_i$ denotes the zero element of $G_i$.
\end{itemize}

Then for any subgroup $K \leq G$, there exist \textbf{unique} subgroups $K_i \leq G_i (1\leq i \leq m)$ such that
\begin{enumerate}[label=\textup{(\roman*)}, itemsep=0pt,align=left,labelwidth=1.5em]
    \item $K = K_1 \oplus K_2 \oplus \cdots \oplus K_m$ as an external direct sum;
    \item $K^\perp =K_1^{\perp} \oplus K_2^{\perp}\oplus \cdots \oplus K_m^{\perp}$ as an external direct sum.
\end{enumerate}
\end{thm}

\begin{proof}

(i) This follows directly from Theorem~\ref{Subgroup Decomposition via Direct Products}, since the orders $|G_i| = |\widetilde{G}_i|$ are pairwise coprime (being prime powers of distinct primes).

(ii) Given the decomposition $G = \mathop{\oplus}\limits_{i=1}^m G_i = \mathop{\oplus}\limits_{i=1}^{m}(\mathop{\oplus}\limits_{j = 1}^{r_i} \mathbb{Z}_{p_i^{\alpha_{ij}}})$, 
we recall the bilinear form in Definition~\ref{shuangxianxingxingshi}:
\[
\langle x, y \rangle := \sum_{i=1}^m \langle x_i, y_i \rangle_i \pmod 1,
\]
where $x_i, y_i \in G_i$, and $\langle x_i, y_i \rangle_i:=\displaystyle\sum_{j=1}^{r_i} \dfrac{x_{ij}y_{ij}}{p_i^{\alpha_{ij}}}\pmod{1}$ with $x_{ij}, y_{ij} \in \mathbb{Z}_{p_i^{\alpha_{ij}}}$.

Following Definition~\ref{zhengjiaoqun}, the subgroup $K^\perp \leq G$ is given by:
\[
K^\perp := \left\{ g \in G \ \middle|\ \langle k, g \rangle = 0 \pmod 1 \text{ for all } k \in K \right\}.
\]
Similarly, for each component $G_i$, we define the subgroup $K_i^\perp \leq G_i$:
\[
K_i^\perp := \left\{ g_i \in G_i \ \middle|\ \langle k_i, g_i \rangle_i = 0 \pmod 1 \text{ for all } k_i \in K_i \right\}.
\]

We establish the equality by proving two inclusions.

\textbf{Step 1: $\mathop{\oplus}\limits_{i=1}^m K_i^\perp \subseteq K^\perp$}

Let $g = (g_1, \ldots, g_m)\in\mathop{\oplus}\limits_{i=1}^m K_i^\perp$ and $k = (k_1, \ldots, k_m) \in \mathop{\oplus}\limits_{i=1}^m K_i$. 
By definition of $K_i^\perp$, we have $\langle k_i, g_i \rangle_i = 0 \pmod 1$ for all $1 \leq i \leq m$. 
Then $\langle k, g \rangle:= \sum\limits_{i=1}^m \langle k_i, g_i \rangle_i \pmod{1} = \sum\limits_{i=1}^m 0 \pmod 1= 0$.

Thus $g \in (\mathop{\oplus}\limits_{i=1}^m K_i)^\perp = K^\perp$, where the equality follows from Part (i). Therefore $\mathop{\oplus}\limits_{i=1}^m K_i^\perp \subseteq K^\perp$.

\textbf{Step 2: $K^\perp \subseteq \mathop{\oplus}\limits_{i=1}^m K_i^\perp$}

Let $g = (g_1, \ldots, g_m) \in K^\perp$. For each $i$ and any $k_i \in K_i$, consider the element
\[
k = (0_1, \ldots, 0_{i-1}, k_i, 0_{i+1}, \ldots, 0_m) \in K,
\]
where the inclusion $k \in K$ follows from Part (i), which gives $K = \mathop{\oplus}\limits_{i=1}^m K_i$.

Since $g \in K^\perp$, we have $\langle k, g \rangle = 0 \pmod 1$. Computing this product:
\[
0 = \langle k, g \rangle = \langle k_i, g_i \rangle_i \pmod 1.
\]
These relations are equivalent for every $k_i \in K_i$, hence we have the equivalence chain:
\[
\langle k_i, g_i \rangle_i = 0 \pmod 1 \ ,\forall k_i \in K_i \ \Longleftrightarrow\  g_i \in K_i^\perp \ \Longleftrightarrow\  g = (g_1, \ldots, g_m) \in \mathop{\oplus}\limits_{i=1}^m K_i^\perp.
\]

Therefore, $K^\perp \subseteq \mathop{\oplus}\limits_{i=1}^m K_i^\perp$. 

Combining Steps 1 and 2, we conclude $K^\perp = \mathop{\oplus}\limits_{i=1}^m K_i^\perp$.
\end{proof}

\begin{remark}
Part (i) of Theorem~\ref{External Direct Sum Decomposition of Subgroups} establishes the theoretical foundation for classical distributed algorithms, while the combination of part (i) and (ii) ensures the correctness of quantum distributed algorithms for finite AHSP in this work.
\end{remark}

\subsection{Division of subfunctions}

The goal of distributed quantum algorithms is to reduce both the qudit numbers and circuit depths at each node, while also minimizing quantum communication complexity. In this subsection, we first introduce the subfunctions and sub-oracles used in Algorithm~\ref{DKL}, then describe finite AHSP in a distributed scenario.  Let finite Abelian group \(G = \mathop{\oplus}\limits_{i=1}^m G_i\), where each \(G_i\) is decomposed as \(G_i = \mathop{\oplus}\limits_{j=1}^{r_i} \mathbb{Z}_{p_i^{\alpha_{ij}}}\).

\begin{defi}\label{zihanshudingyi}
The original black-box function is $f: G \to G$ with $G=\mathop{\oplus}\limits_{i=1}^m G_i$, where $G_i$ is the Sylow-$p_i$ subgroups of $G$. We define $m$ sub-functions $\{f_i\}_{i=1}^m: G_i\to G$, each being a restriction of $f$ to $G_i$: 
\[
f_i(g_i) = f\bigl(0_1,\, \dots,\, 0_{i-1},g_i, 0_{i+1} \dots,\, 0_m\bigr), \quad \forall g_i \in G_i.
\]
Each sub-function $f_i$ acts exclusively on the $i$-th projection coordinate. 
\end{defi}

In the centralized AHSP, a single oracle $U_f$ has access to the whole function. For any $g \in G$, it acts as
\[
U_{f}\ket{g}\ket{b} = \ket{g}\ket{b + f(g)}, \quad b \in G.
\]

By contrast, our distributed quantum algorithm involves $m$ parties, where each party $\mathcal{P}_i$ is equipped with a local oracle $U_{f_i}$ that is defined only on the subgroup $G_i$. Specifically, for any $g_i \in G_i$, the oracle acts as
\begin{equation}\label{zioracle}
U_{f_i}\ket{g_i}\ket{b} = \ket{g_i}\ket{b + f_i(g_i)}, \quad b \in G.
\end{equation}

This framework is particularly suitable for multipartite quantum systems where each participant has only partial access to the function. In such systems, the parties collaboratively determine the hidden subgroup $K$ through local oracle queries. Combining Theorem~\ref{External Direct Sum Decomposition of Subgroups} with the sub-functions $f_i$ in Definition~\ref{zihanshudingyi}, we obtain Theorem~\ref{fenbushidingli}.

\begin{thm}\label{fenbushidingli}
 Let function $f: G\rightarrow G$ hide the subgroup $K$, i.e., $f(x) = f(y)\Longleftrightarrow x-y\in K$, where $K\leq G$. Let $K=\mathop{\oplus}\limits_{i=1}^m K_i$, where $K_i$ is defined in Theorem~\ref{External Direct Sum Decomposition of Subgroups}(i). Then for all $x_i, y_i \in G_i$, where $G_i$ is the Sylow $p_i$-subgroup of $G$, the function $f_i$ hides the subgroup $K_i$, i.e., $f_i(x_i)=f_i(y_i)\Longleftrightarrow x_i-y_i\in K_i$.
\end{thm}

\begin{proof}
\begin{align}
f_i(x_i) = f_i(y_i) 
    &\Longleftrightarrow f(0_1,\dots,x_i, \dots,0_m)=f(0_1,\dots,y_i, \dots,0_m) \quad\text{(by Definition~\ref{zihanshudingyi})}\nonumber \\
    &\Longleftrightarrow (0_1,\dots,x_i, \dots,0_m)-(0_1,\dots,y_i, \dots,0_m)\in K \nonumber \\
    & \hspace{2cm} \text{(by property of $f$: $f(x) = f(y) \Leftrightarrow x-y\in K$)}\nonumber \\
    &\Longleftrightarrow (0_1,\dots,x_i-y_i, \dots,0_m)\in K \nonumber
\end{align}

By Theorem~\ref{External Direct Sum Decomposition of Subgroups}(i), $K=\mathop{\oplus}\limits_{i=1}^m K_i$. Thus
\begin{align*}
(0_1,\dots,x_i-y_i, \dots,0_m)\in K\Longleftrightarrow x_i-y_i\in K_i. 
\end{align*}
\end{proof}

Define $\mathrm{QFT}_{G_i}:=\mathop{\textstyle\bigotimes}\limits_{j=1}^{r_i} \mathrm{QFT}_{p_i^{\alpha_{ij}}}$ and $\mathrm{QFT}_{G_i}^\dagger:=\mathop{\textstyle\bigotimes}\limits_{j=1}^{r_i} \mathrm{QFT}_{p_i^{\alpha_{ij}}}^\dagger$, then we have Lemma~\ref{QFTdagger}.

\begin{lemma}\label{QFTdagger} 
For any $\bm{u_i},\bm{m_i}\in G_i$, we have
\[
\mathrm{QFT}_{G_i}^\dagger|\bm{u_i}\rangle = \frac{1}{\sqrt{|G_i|}} 
\sum_{\bm{m_i}\in G_i} e^{-2\pi i \langle\bm{u_i}, \bm{m_i}\rangle_i}|\bm{m_i}\rangle,
\]
where $\langle \bm{u_i}, \bm{m_i} \rangle_i :=\displaystyle\sum_{j=1}^{r_i} \dfrac{u_{ij}m_{ij}}{p_i^{\alpha_{ij}}}\bmod{1}$, with $u_{ij}, m_{ij} \in \mathbb{Z}_{p_i^{\alpha_{ij}}}$.
\end{lemma}

\begin{proof}
\begin{align*}
\mathrm{QFT}_{G_i}^\dagger|\bm{u_i}\rangle=&\left(\mathrm{QFT}_{p_i^{\alpha_{i1}}}^\dagger \otimes \cdots \otimes \mathrm{QFT}_{p_i^{\alpha_{ir_i}}}^\dagger \right) (\Ket{u_{i1}} \Ket{u_{i2}} \cdots \Ket{u_{iri}})\\
=&({\rm QFT}_{p_i^{\alpha_{i1}}}^{\dagger}\Ket{u_{i1}})\otimes ({\rm QFT}_{p_i^{\alpha_{i2}}}^{\dagger}\Ket{u_{i2}})\otimes\cdots\otimes ({\rm QFT}_{p_i^{\alpha_{ir_i}}}^{\dagger}\Ket{u_{ir_i}})\\
=&(\frac{1}{\sqrt{p_i^{\alpha_{i1}}}}\sum\limits_{m_{i1}=0}^{p_i^{\alpha_{i1}}-1} e^{-2\pi iu_{i1}\cdot m_{i1}/p_i^{\alpha_{i1}}}|m_{i1}\rangle)\otimes\cdots\otimes (\frac{1}{\sqrt{p_i^{\alpha_{ir_i}}}}\sum\limits_{m_{ir_i}=0}^{p_i^{\alpha_{ir_i}}-1} e^{-2\pi iu_{ir_i}\cdot m_{ir_i}/p_i^{\alpha_{ir_i}}}|m_{ir_i}\rangle)\\
=&\frac{1}{\sqrt{|G_i|}}\sum_{\bm{m_i}\in G_i}e^{-2\pi i\langle\bm{u_i}, \bm{m_i}\rangle_i}|\bm{m_i}\rangle.
\end{align*}
\end{proof}

\subsection{Distributed exact quantum algorithm for finite AHSP}

\begin{figure*}[htbp]
  \centering
  \includegraphics[scale=0.4]{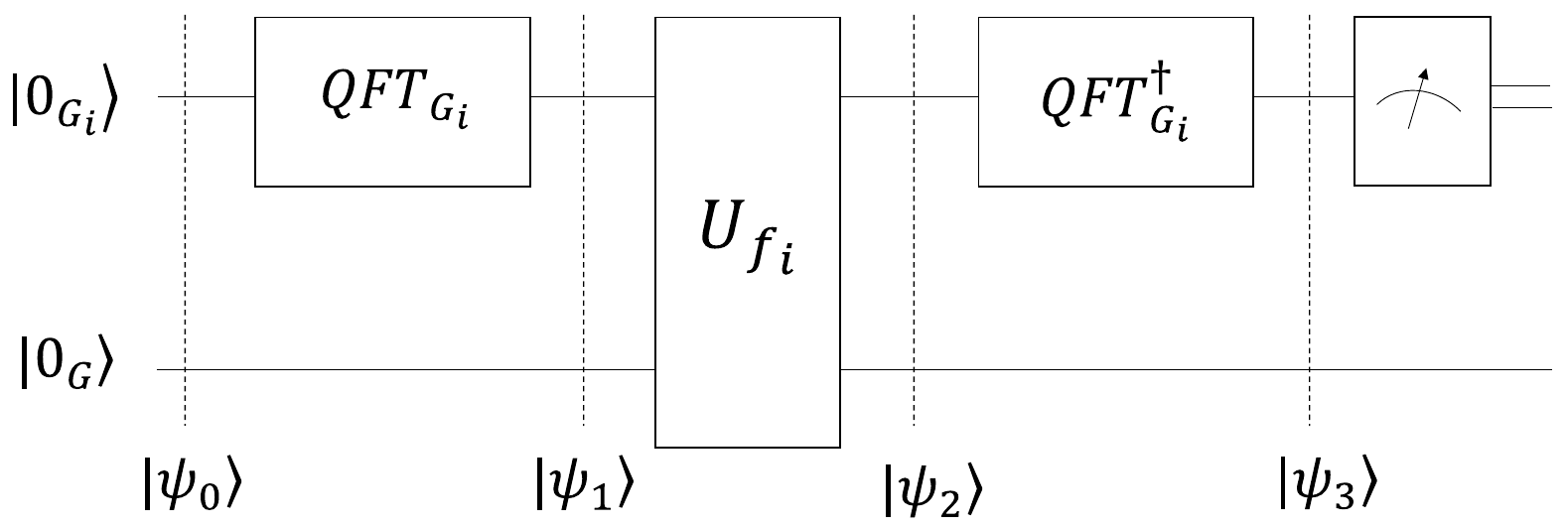}
  \caption{The circuit for DKL in the $i$-th node (Algorithm~\ref{DKL}).}
  \label{algorithm4_circuit}
\end{figure*}

In the following, we present our distributed quantum algorithm (Algorithm~\ref{DKL}) for determining $K_i$, along with its quantum circuit implementation in Fig.~\ref{algorithm4_circuit}. In line 4 of Algorithm~\ref{DKL} and Fig.~\ref{algorithm4_circuit}, we define $0_{G_i}:=(0_{i1}, 0_{i2},\dots, 0_{ir_i})$ as the zero element of group $G_i$, and $0_G:=(0_{G_1}, 0_{G_2},\dots, 0_{G_m})$ as the zero element of group $G$.

In line 4, $\mathcal{H}_{G_i}(1\le i\le m)$ is a $|G_i|$-dimensional Hilbert space with basis $\{\lvert x_i \rangle : x_i =0, 1, \ldots, |G_i|-1\}$, and $\mathcal{H}_G=\mathop{\textstyle\bigotimes}\limits_{i=1}^m \mathcal{H}_{G_i}$ is their tensor product. In lines 5 and 7, define $\mathrm{I}_{G}:=\mathop{\textstyle\bigotimes}\limits_{i=1}^m \mathrm{I}_{G_i}$, where $\mathrm{I}_{G_i}$ is the identity operator on $\mathcal{H}_{G_i}$; the quantum Fourier transform and its inverse are defined as $\mathrm{QFT}_{G_i}:=\mathop{\textstyle\bigotimes}\limits_{j=1}^{r_i} \mathrm{QFT}_{p_i^{\alpha_{ij}}}$ and $\mathrm{QFT}_{G_i}^\dagger:=\mathop{\textstyle\bigotimes}\limits_{j=1}^{r_i} \mathrm{QFT}_{p_i^{\alpha_{ij}}}^\dagger$. In line~6, the oracle $U_{f_i}$ is applied as defined in Eq.~(\ref{zioracle}).

\begin{algorithm}[t]
  \caption{Local quantum algorithm for $K_i$ at node $i$}\label{DKL}
  \begin{algorithmic}[1]
    \Statex \textbf{Input:} Group $G_i$, oracle $U_{f_i}$, success probability $1-\epsilon/m$
    \Statex \textbf{Output:} Subgroup $A_i$ containing hidden subgroup $K_i$
    
    \State $M_0 \gets \emptyset$;
    \State $h \gets \min \left\{ \mathrm{rank}(G_i)+\lceil\log_2(2m/\epsilon)\rceil, \ \mathrm{len}(G_i)+\lceil\log_2(m/\epsilon)\rceil \right\}$; \Comment{By Theorem~\ref{thm8}} 
    \For{$j = 1$ to $h$}
 \State $|\psi_0\rangle = |0_{G_i}\rangle|0_G\rangle \in \mathcal{H}_{G_i} \otimes \mathcal{H}_{G}$;
      \State $|\psi_1\rangle = \bigg(\mathrm{QFT}_{G_i}\otimes \mathrm{I}_{G}\bigg)|\psi_0\rangle$;
      \State $|\psi_2\rangle = U_{f_i}|\psi_1\rangle$;
      \State $|\psi_3\rangle = \bigg(\mathrm{QFT}_{G_i}^\dagger\otimes \mathrm{I}_{G}\bigg)|\psi_2\rangle$;
      \State Measure the first register to obtain $\bm{m}_j = (m_{j1}, m_{j2},\dots, m_{jr_i}) \in K_i^{\perp}$;
      \State Update $M_j \gets M_{j-1} \cup \{\bm{m}_j\}$.
    \EndFor
    
    \State Solve linear congruence system:
    \[
    \bm{W}\bm{x}^{\mathrm{T}} \equiv \bm{0}^{\mathrm{T}} \pmod{1} \quad \text{where } \bm{W} = \begin{pmatrix} 
    m_{11}/p_i^{\alpha_{i1}} & m_{12}/p_i^{\alpha_{i2}} & \cdots & m_{1r_i}/p_i^{\alpha_{i r_i}} \\
    \vdots & \vdots & \ddots & \vdots \\
    m_{h1}/p_i^{\alpha_{i1}} & m_{h2}/p_i^{\alpha_{i2}} & \cdots & m_{hr_i}/p_i^{\alpha_{i r_i}}
    \end{pmatrix}
    \]
    
    \State Solution subgroup: $A_i = (\mathrm{span}(M_h))^{\perp} \supseteq K_i^{\perp\perp} = K_i$;
    \State \Return $A_i$
  \end{algorithmic}
\end{algorithm}

Before proving the correctness of Algorithm~\ref{DKL}, we first give Lemma~\ref{Lemma3}.

\begin{lemma}\label{Lemma3}
Let $\bm{m} \notin K_i^{\perp}$. Then the following holds:
\[
\sum_{\bm{k_i} \in K_i} e^{-2\pi i \langle \bm{k_i}, \bm{m} \rangle_i} = 0.
\]
\end{lemma}

\begin{proof}
Since $\bm{m} \notin K_i^{\perp}$, there exists some $\bm{k'_i} \in K_i$ such that $\langle \bm{k'_i}, \bm{m} \rangle_i \not\equiv 0 \pmod{1}$. The additive closure of $K_i$ implies that for every $\bm{k_i} \in K_i$, we have $\bm{k_i} + \bm{k'_i} \in K_i$.

Consider the sum:
\begin{equation}\label{shizi21}
\sum_{\bm{k_i} \in K_i} e^{-2\pi i \langle \bm{k_i} + \bm{k'_i}, \bm{m} \rangle_i}
= e^{-2\pi i \langle \bm{k'_i}, \bm{m} \rangle_i} \sum_{\bm{k_i} \in K_i} e^{-2\pi i \langle \bm{k_i}, \bm{m} \rangle_i}.
\end{equation}

Let $C=\displaystyle\sum_{\bm{k_i}\in K_i}^{} e^{-2\pi i\langle \bm{k_i},\bm{m}\rangle_i}$. Since the map $\bm{k_i} \mapsto \bm{k_i} + \bm{k'_i}$ is a bijection on $K_i$, the left-hand side of Eq.~(\ref{shizi21}) equals $C$. Thus we obtain:
\[
C = e^{-2\pi i \langle \bm{k'_i}, \bm{m} \rangle_i} C
\quad\Longleftrightarrow \quad 
C \left(1 - e^{-2\pi i \langle \bm{k'_i}, \bm{m} \rangle_i}\right) = 0.
\]

Since $\langle \bm{k'_i}, \bm{m} \rangle_i \not\equiv 0 \pmod{1}$, it follows that $e^{-2\pi i \langle \bm{k'_i}, \bm{m} \rangle_i} \neq 1$. Therefore $C = 0$, which completes the proof.
\end{proof}

Next we give Theorem~\ref{thm6}.

\begin{thm}\label{thm6}
In line 8 of Algorithm~\ref{DKL}, the measurement result satisfies
$\bm{m}_j=(m_{j1}, \dots, m_{jr_i})\in K_i^\perp$ with certainty.
\end{thm}
\begin{proof}

In line 5, we use quantum Fourier transform to get a uniform superposition state in high-dimensional Hilbert space. The state after line 5 of Algorithm~\ref{DKL} is

\begin{align*}
 |\psi_1\rangle&=\bigg(\mathrm{QFT}_{G_i}\otimes \mathrm{I}_{G}\bigg)|\psi_0\rangle\\
&=\left(\mathrm{QFT}_{p_i^{\alpha_{i1}}} \otimes \cdots \otimes \mathrm{QFT}_{p_i^{\alpha_{ir_i}}}\otimes \mathrm{I}_{G}\right) (\overbrace{|0\rangle|0\rangle \ldots|0\rangle}^{r_i}|0_G\rangle)\\
&=\frac{1}{\sqrt{|G_i|}} 
\sum_{\substack{(u_1, u_2, \cdots, u_{r_i}) \in G_i}}\Ket{u_1} \Ket{u_2} \cdots \Ket{u_{r_i}}|0_G\rangle\\
&=\frac{1}{\sqrt{|G_i|}} 
\sum_{\bm{u_i}\in G_i}\Ket{\bm{u_i}}|0_G\rangle).
\end{align*}

In line 6 of Algorithm~\ref{DKL}, we have the state
\begin{align*}
|\psi_2\rangle=&U_{f_i}|\psi_1\rangle\\
=&\frac{1}{\sqrt{|G_i|}} 
\sum_{\bm{u_i}\in G_i}\Ket{\bm{u_i}}\Ket{f_i(\bm{u_i})}.
\end{align*}

In line 7 of Algorithm~\ref{DKL}, we use high-dimensional inverse quantum Fourier transform on the first register,
\begin{align}
	|\psi_3\rangle&=\bigg(\mathrm{QFT}_{G_i}^\dagger\otimes \mathrm{I}_{G}\bigg)|\psi_2\rangle\nonumber\\
&=\bigg(\mathrm{QFT}_{G_i}^\dagger\otimes \mathrm{I}_{G}\bigg)\frac{1}{\sqrt{|G_i|}} 
\sum_{\bm{u_i}\in G_i}\Ket{\bm{u_i}}\Ket{f_i(\bm{u_i})}\nonumber\\
&=\frac{1}{\sqrt{|G_i|}}\sum_{\bm{u_i}\in G_i}\left(\mathrm{QFT}_{G_i}^\dagger|\bm{u_i}\rangle\right)\Ket{f_i(\bm{u_i})}\nonumber\\
&=\frac{1}{|G_i|}\sum_{\bm{u_i}\in G_i}\sum_{\bm{m}\in G_i}e^{-2\pi i\langle\bm{u_i}, \bm{m}\rangle_i}|\bm{m}\rangle\Ket{f_i(\bm{u_i})}\label{142shizi}\\
&=\frac{1}{|G_i||K_i|}\sum_{\bm{u_i}\in G_i}\sum_{\bm{m}\in G_i}\sum_{\bm{k_i}\in K_i}e^{-2\pi i\langle\bm{u_i}+\bm{k_i}, \bm{m}\rangle_i}|\bm{m}\rangle\Ket{f_i(\bm{u_i}+\bm{k_i})}\nonumber\\
&=\frac{1}{|G_i||K_i|}\sum_{\bm{u_i}\in G_i}\sum_{\bm{m}\in G_i}\sum_{\bm{k_i}\in K_i}e^{-2\pi i\langle\bm{u_i}+\bm{k_i}, \bm{m}\rangle_i}|\bm{m}\rangle\Ket{f_i(\bm{u_i})}\label{144shizi}\\
&=\frac{1}{|G_i||K_i|} \sum_{\bm{u_i} \in G_i} \sum_{\bm{m} \in K_i^\perp} 
\left( \sum_{\bm{k_i} \in K_i} e^{-2\pi i \langle \bm{k_i}, \bm{m} \rangle_i} \right) 
e^{-2\pi i \langle \bm{u_i}, \bm{m} \rangle_i} |\bm{m}\rangle 
\Ket{f_i(\bm{u_i})} \nonumber\\
& + \frac{1}{|G_i||K_i|} \sum_{\bm{u_i} \in G_i} \sum_{\bm{m} \in G \setminus K_i^\perp} 
\left( \sum_{\bm{k_i} \in K_i} e^{-2\pi i \langle \bm{k_i}, \bm{m} \rangle_i} \right) 
e^{-2\pi i \langle \bm{u_i}, \bm{m} \rangle_i} |\bm{m}\rangle \Ket{f_i(\bm{u_i})}\nonumber\\
&=\frac{1}{|G_i||K_i|} \sum_{\bm{u_i} \in G_i} \sum_{\bm{m} \in K_i^\perp} 
\left( \sum_{\bm{k_i} \in K_i} e^{-2\pi i \langle \bm{k_i}, \bm{m} \rangle_i} \right) 
e^{-2\pi i \langle \bm{u_i}, \bm{m} \rangle_i} |\bm{m}\rangle 
\Ket{f_i(\bm{u_i})}\label{147shizi}\\
&=\frac{1}{|G_i|} \sum_{\bm{u_i} \in G_i} \sum_{\bm{m} \in K_i^\perp}e^{-2\pi i \langle \bm{u_i}, \bm{m} \rangle_i} |\bm{m}\rangle \Ket{f_i(\bm{u_i})}.\nonumber
\end{align}

Eq.~(\ref{142shizi}) follows from Lemma~\ref{QFTdagger}, while Eq.~(\ref{144shizi}) results from $f_i(\bm{u_i})=f_i(\bm{u_i}+\bm{k_i})$ in Theorem~\ref{fenbushidingli}, and Eq.~(\ref{147shizi}) is given by Lemma~\ref{Lemma3}. Thus in line 8 of Algorithm~\ref{DKL}, measuring the first register, we get the a string of $\bm{m}_j\in K_i^{\perp}$ with certainty. 
\end{proof}

Therefore, in line 12 of Algorithm~\ref{DKL}, we obtain a subgroup $A_i=\mathrm{span}(M_h)^{\perp}\supseteq K_i^{\perp\perp} = K_i$, and we proceed to analyze the success probability of Algorithm~\ref{DKL} in Theorem~\ref{thm8}.

\begin{thm}\label{thm8}
In Algorithm~\ref{DKL}, to ensure $\Pr(A_i = K_i) \geq 1 - \dfrac{\epsilon}{m}$, it is sufficient to set
\[
h \ge \mathrm{rank}(G_i) + \left\lceil \log_2 \dfrac{2m}{\epsilon} \right\rceil \quad \text{or} \quad h \ge \mathrm{len}(G_i) - \mathrm{len}(K_i) + \left\lceil \log_2 \dfrac{m}{\epsilon} \right\rceil.
\]
\end{thm}

\begin{proof}

It follows directly from Proposition~\ref{zhihuolianchangdingli}. Note that ensuring \(\Pr(A_i = K_i) \ge 1 -  \dfrac{\epsilon}{m}\) is equivalent to ensuring \(\Pr(\langle \mathbf{m}_1, \mathbf{m}_2, \dots, \mathbf{m}_h \rangle = K_i^\perp) = \varphi_h(K_i^\perp) \ge 1-\dfrac{\epsilon}{m}\). By Proposition~\ref{zhihuolianchangdingli}, this probability holds if we choose:
\[
h \ge \mathrm{rank}(K_i^\perp) + \left\lceil \log_2 \frac{2m}{\epsilon} \right\rceil \quad \text{or} \quad h \ge \mathrm{len}(K_i^\perp) + \left\lceil \log_2 \frac{m}{\epsilon} \right\rceil.
\]
Since $\mathrm{rank}(K_i^\perp)\le\mathrm{rank}(G_i)$ and $\mathrm{len}(K_i^\perp) = \mathrm{len}(G_i) - \mathrm{len}(K_i)$, 
we obtain the required bounds:

\[
h \ge \mathrm{rank}(G_i) + \left\lceil \log_2 \frac{2m}{\epsilon} \right\rceil \quad \text{or} \quad h \ge \mathrm{len}(G_i) - \mathrm{len}(K_i) + \left\lceil \log_2 \frac{m}{\epsilon} \right\rceil.
\]
 
\end{proof}

\begin{thm}\label{thm9}
Consider Algorithm~\ref{DKL} over $m$ nodes. Let $A = \mathop{\oplus}\limits_{i=1}^m A_i$ be the composite subgroup, with $K = \mathop{\oplus}\limits_{i=1}^m K_i$ as the target subgroup. If the iteration number $h$ at each node is chosen large enough so that $\Pr(A_i = K_i) \geq 1 - \epsilon/m$ (as guaranteed by Theorem~\ref{thm8}), then
\[
\Pr(A = K) \geq 1 - \epsilon.
\]
\end{thm}

\begin{proof}
By Theorem~\ref{thm8}, our choice of $h$ ensures $\Pr(A_i = K_i) \geq 1 - \epsilon/m$ for each node $i\in\{1, \dots, m\}$. Since $A = \mathop{\oplus}\limits_{i=1}^m A_i$ and $K =\mathop{\oplus}\limits_{i=1}^m K_i$, we have $A = K$ if and only if $A_i = K_i$ for all $i$. 

Given the mutual independence of the events $\{A_i = K_i\}_{i=1}^m$, it follows that:
\begin{align*}
\Pr(A = K) &= \prod_{i=1}^m \Pr(A_i = K_i) \\
&\geq \prod_{i=1}^m \left(1 - \frac{\epsilon}{m}\right) \\
&\geq 1 - m \cdot \frac{\epsilon}{m} \quad \text{(Bernoulli's inequality)} \\
&= 1 - \epsilon.
\end{align*}
\end{proof}

When $|K|$ is known, for any $1\le i\le m$, the subgroup order $|K_i|$ is determined by
\[
|K_i| = \gcd(|K|,|G_i|).
\]

Given the order $|K|$, Algorithm~\ref{EDKL} achieves $\Pr(A_i=K_i)=1$ via Algorithm~\ref{algorithm5} as a quantum subroutine. Algorithm~\ref{algorithm5} extends the quantum amplitude amplification (QAA) framework of Algorithm~\ref{algorithm2} by substituting  $(G, K, f) \to (G_i, K_i, f_i)$. Let \( \Omega = \{\bm{u}_1, \dotsc,\bm{u}_{[G_i:K_i]}\} \) be coset representatives for \( K_i \leq G_i \). Define the composite unitary operation from lines 5 to 7 in Algorithm~\ref{DKL}:
\[
\mathcal{A}=(\mathrm{QFT}_{G_i}^\dagger \otimes \mathrm{I}_{G})\cdot U_{f_i}\cdot (\mathrm{QFT}_{G_i} \otimes \mathrm{I}_{G}).
\]
Let $|\psi_3\rangle$ denote the state after line 7 in Algorithm~\ref{DKL}. In the $j$-th iteration, we define the phase rotation operators $\mathcal{R}_{0}(\phi_j)$ and $\mathcal{R}_{\mathcal{A}}(\varphi_j, M_{j-1})$ as follows:

\begin{equation*}
\mathcal{R}_0(\phi_j)\Ket{a,b} = 
\begin{cases}
    \Ket{a,b}, & (a,b) \neq (0_{G_i}, 0_G) \in G_i \oplus G \\
    e^{\imath\phi_j}\Ket{a,b}, & (a,b) = (0_{G_i}, 0_G) \in G_i \oplus G 
\end{cases}
\end{equation*}
where $\imath = \sqrt{-1}$, and
\begin{equation*}
\mathcal{R}_{\mathcal{A}}(\varphi_j, M_{j-1})\Ket{a} = 
\begin{cases}
    e^{\imath\varphi_j}\Ket{a}, & a \notin \mathrm{span}(M_{j-1}) \\
    \Ket{a}, & a \in \mathrm{span}(M_{j-1}).
\end{cases}
\end{equation*}

Using these operators, we construct the quantum amplitude amplification operator:
\[
\mathcal{Q}_j = \mathcal{A} \mathcal{R}_{0}(\phi_j) \mathcal{A}^{\dagger} \left(\mathcal{R}_{\mathcal{A}}(\varphi_j, M_{j-1}) \otimes \mathrm{I}_{G}\right).
\]

Let $Z_j = K_i^{\perp} \setminus \mathrm{span}(M_{j-1})$. We define the good subsets and bad subsets:
\begin{itemize}
    \item $\Ket{\Psi_{Z_j}}$: good subset spanned by $\{\Ket{a,b} \mid a \in Z_j, b \in G\}$ 
    \item $\Ket{\Psi_{Y_j}}$: bad subset spanned by $\{\Ket{a,b} \mid a \in \mathrm{span}(M_{j-1}), b \in G\}$
\end{itemize}
Note that $\Ket{\Psi_{Z_j}}$ and $\Ket{\Psi_{Y_j}}$ are not quantum states, as they are not normalized.

The state of $|\psi_3\rangle$ decomposes as:
\[
|\psi_3\rangle = \Ket{\Psi_{Z_j}} + \Ket{\Psi_{Y_j}}.
\]

\begin{algorithm}[t]
  \caption{Quantum amplitude amplification for measuring good states (QAA)}\label{algorithm5}
  \begin{algorithmic}[1]
    \Statex \textbf{Input:} Registers $|\psi_3\rangle$, operator $\mathcal{A}$, set $M_{j-1}$
    \Statex \textbf{Output:} $\bm{m}_j \in K_i^{\perp} \setminus \mathrm{span}(M_{j-1})$
    
    \State $b_j \gets 1 - \dfrac{|\mathrm{span}(M_{j-1})| \cdot |K_i|}{|G_i|}$;
    \State $\phi_j \gets \arccos\left(1 - \dfrac{1}{2b_j}\right)$ or $2\pi - \arccos\left(1 - \dfrac{1}{2b_j}\right)$;
    \State $\varphi_j \gets \arccos\left(1 - \dfrac{1}{2b_j}\right)$ or $2\pi - \arccos\left(1 - \dfrac{1}{2b_j}\right)$;
    \State Apply $\mathcal{Q}_j$ once to $|\psi_3\rangle$ to obtain $|\psi_4\rangle = \mathcal{Q}_j|\psi_3\rangle$, where $\mathcal{Q}_j = \mathcal{A}\mathcal{R}_{0}(\phi_j)\mathcal{A}^{\dagger}\left(\mathcal{R}_{\mathcal{A}}(\varphi_j, M_{j-1})\otimes \mathrm{I}_{G}\right)$;
    \State Measure the first register to obtain $\bm{m}_j = (m_{j1}, m_{j2}, \dots, m_{jr_i}) \in K_i^{\perp} \setminus \mathrm{span}(M_{j-1})$;
    \State \Return $\bm{m}_j$
  \end{algorithmic}
\end{algorithm}

\begin{thm}\label{jingquefenbushiziguochengdingli} 
Let $\varphi_j=\phi_j=\arccos(1-\dfrac{1}{2b_j})$ or $\varphi_j=\phi_j=2\pi-\arccos(1-\dfrac{1}{2b_j})$, where $b_j = 1 - \dfrac{|\mathrm{span}(M_{j-1})| \cdot |K_i|}{|G_i|}$. Then $\mathcal{Q}_j|\psi_3\rangle =\Ket{\Psi_{Z_j}}$.

Hence, line 5 of Algorithm~\ref{algorithm5} deterministically yields $\bm{m}_j\notin \mathrm{span}(M_{j-1})$.
\end{thm}

\begin{proof}
The proof is analogous to Theorem~\ref{jingqueziguochengdingli} by substituting  $(G, K, f)$ to $(G_i, K_i, f_i)$.
\end{proof}

\begin{figure*}[htbp]
  \centering
  \includegraphics[scale=0.4]{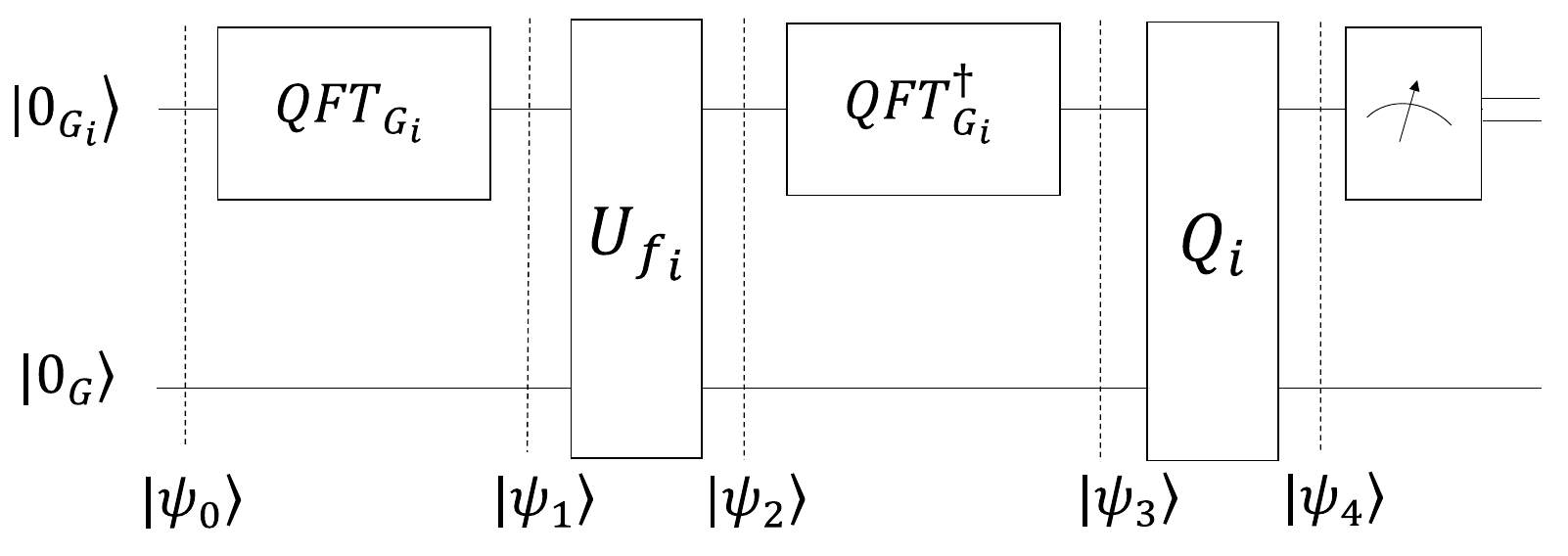}
  \caption{The circuit for EDKL in the $i$-th node (Algorithm~\ref{EDKL}).}
  \label{algorithm6_circuit}
\end{figure*}

Combining Algorithm~\ref{DKL} with Algorithm~\ref{algorithm5}, we obtain the exact distributed quantum algorithm for Abelian subgroup problem, formally presented in Algorithm~\ref{EDKL}. The quantum circuit of Algorithm~\ref{EDKL} is shown in Fig.~\ref{algorithm6_circuit}. In Algorithm~\ref{EDKL} and Fig.~\ref{algorithm6_circuit}, we define $0_{G_i}:=(0_{i1}, 0_{i2},\dots, 0_{ir_i})$ as the zero element of group $G_i$, and $0_G:=(0_{G_1}, 0_{G_2},\dots, 0_{G_m})$ as the zero element of group $G$.

\begin{algorithm}[h]
  \caption{Local exact quantum algorithm for $K_i$ at node $i$ (EDKL)}\label{EDKL}
  \begin{algorithmic}[1]
    \Statex \textbf{Input:} Group $G_i$, oracle $U_{f_i}$
    \Statex \textbf{Output:} Subgroup $A_i = \operatorname{span}(M_h) = K_i^\perp$

    \Function{EDKL}{$G_i$, $U_{f_i}$}
      \State $M_0 \gets \emptyset$;
      \State $h \gets \operatorname{len}(G_i) - \operatorname{len}(K_i)$;
      \For{$j = 1$ \textbf{to} $h$}
        \State $|\psi_0\rangle=|0_{G_i}\rangle|0_G\rangle \in \mathcal{H}_{G_i} \otimes \mathcal{H}_{G}$;
        \State Apply $\mathcal{A}$ to obtain $|\psi_3\rangle = \mathcal{A}|\psi_0\rangle$, where $\mathcal{A} = (\mathrm{QFT}_{G_i}^\dagger \otimes \mathrm{I}_{G}) \cdot U_{f_i} \cdot (\mathrm{QFT}_{G_i} \otimes \mathrm{I}_{G})$;
        \State $\bm{m}_j \gets$ \Call{QAA}{$|\psi_3\rangle$, $\mathcal{A}$, $M_{j-1}$};
        \State Update $M_j \gets M_{j-1} \cup \{\bm{m}_j\}$; 
      \EndFor
      \State \Return $A_i=\operatorname{span}(M_h)= K_i^\perp$
    \EndFunction
  \end{algorithmic}
\end{algorithm}

\begin{Pp}\label{thm10}
     In Algorithm~\ref{EDKL}, in the $i$-th node$(1\le i\le m)$, after performing $\mathrm{len}(G_i)-\mathrm{len}(K_i)$ iterations, we can exactly obtain $A_i=K_i^\perp$.
\end{Pp}

\begin{proof}
In line 6 of Algorithm~\ref{EDKL}, we get the state $|\psi_3\rangle=\mathcal{A}|\psi_0\rangle$. Applying $\mathcal{Q}_j$ on $|\psi_3\rangle$ yields $|\psi_4\rangle=\mathcal{Q}_j|\psi_3\rangle$, where $\mathcal{Q}_j=\mathcal{A}\mathcal{R}_{0}(\phi_j)\mathcal{A}^{\dagger}\left(\mathcal{R}_{\mathcal{A}}(\varphi_j, M_{j-1})\otimes \mathrm{I}_{G}\right)$, $\mathcal{A}=(\mathrm{QFT}_{G_i}^\dagger \otimes \mathrm{I}_{G})\cdot U_{f_i}\cdot (\mathrm{QFT}_{G_i} \otimes \mathrm{I}_{G})$, $\phi_j=\varphi_j=\arccos(1-\dfrac{1}{2b_j})$ or $\phi_j=\varphi_j=2\pi-\arccos(1-\dfrac{1}{2b_j})$, $b_j=1 - \dfrac{|\mathrm{span}(M_{j-1})|\cdot|K_i|}{|G_i|}$.

Theorem~\ref{jingquefenbushiziguochengdingli} implies that $|\psi_4\rangle=|\Psi_{Z_j}\rangle$. Therefore, measuring the first register of \(|\psi_4\rangle\) in line 7 of Algorithm~\ref{EDKL} deterministically yields $\bm{m}_j\notin \mathrm{span}(M_{j-1})$. Thus after completing all $h = \mathrm{len}(G_i) - \mathrm{len}(K_i)$ iterations, we have $\mathrm{len}(\mathrm{span}(M_h)) \ge \mathrm{len}(K_i^\perp) = \mathrm{len}(G_i) - \mathrm{len}(K_i)$, which implies $A_i=\mathrm{span}(M_h) = K_i^{\perp}$. 

\end{proof}

In our distributed algorithm, a total of $m$ quantum nodes are required. Each node independently finds $K_i^\perp(1\le i\le m)$ , and transmits the obtained \( K_i^\perp \) to the central node via classical communication. According to part (ii) of Theorem~\ref{External Direct Sum Decomposition of Subgroups}, combining the \( K_i^\perp \) from each node yields 
\[
K^\perp = K_1^\perp \oplus K_2^\perp \oplus \cdots \oplus K_m^\perp.
\]

Subsequently, the central node performs classical post-processing (solving a linear congruence system) to compute \( K^{\perp\perp}=K \).

\begin{figure*}[htbp]
  \centering
  \includegraphics[width=\textwidth]{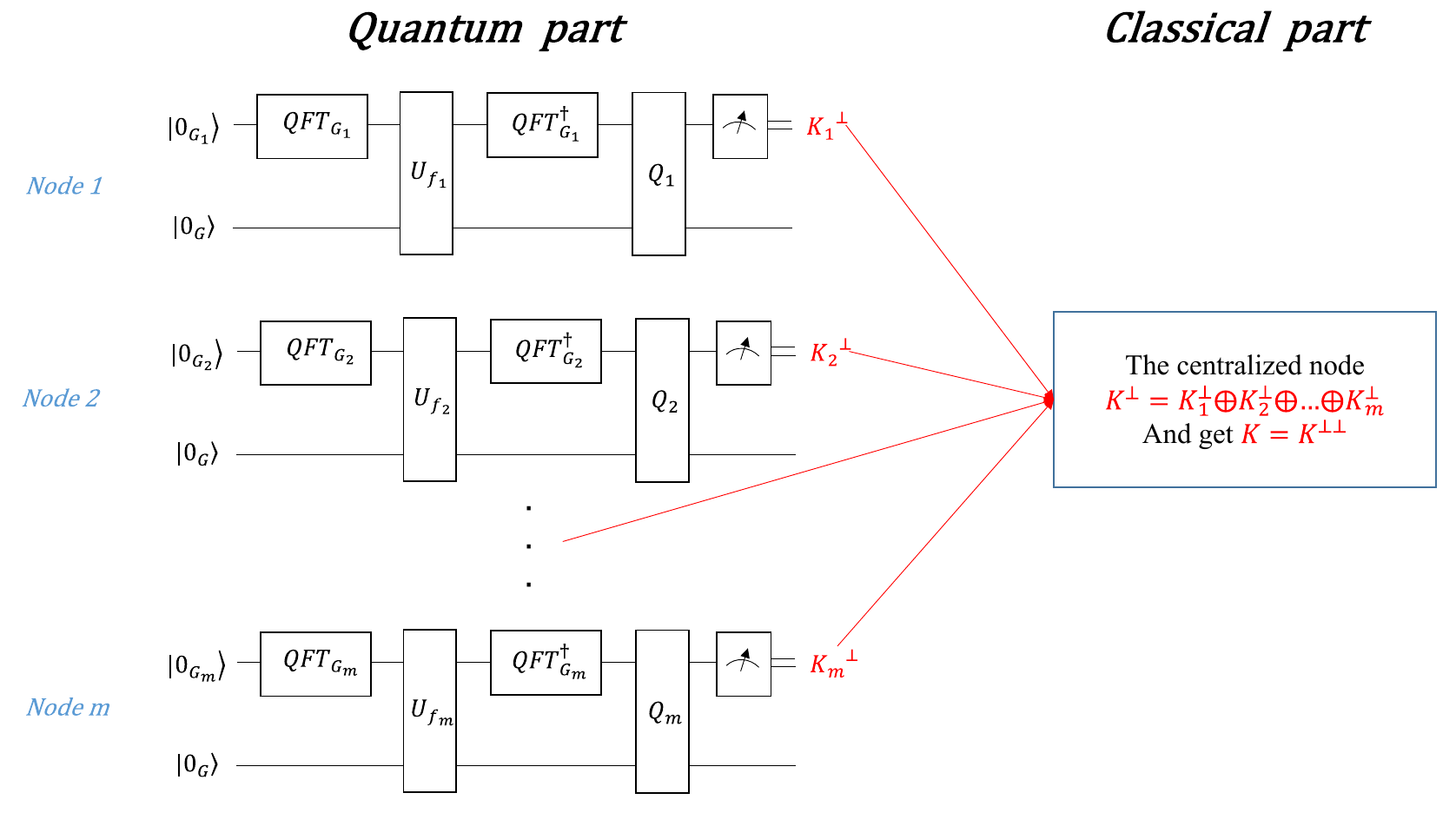}
  \caption{The whole circuit for EDK (Algorithm~\ref{EDK}).}
  \label{algorithm7_circuit}
\end{figure*}

Our Algorithm~\ref{EDK} establishes a pure LOCC (Local Operations and Classical Communication) model: $m$ nodes perform parallel operations using only classical communication, eliminating quantum communication requirements. The whole distributed quantum algorithm is formally presented in Algorithm~\ref{EDK}, with its quantum circuit shown in Fig.~\ref{algorithm7_circuit}. In Fig.~\ref{algorithm7_circuit}, for each node $i$, we define $0_i := (0_{i1}, 0_{i2}, \dots, 0_{ir_i})$ as the zero element of the group $G_i$. Then we present Theorem~\ref{thm14}, which guarantees the correctness of Algorithm~\ref{EDK}.

\begin{algorithm}[h]
  \caption{Exact distributed quantum algorithm for finding the entire subgroup $K$}\label{EDK}
  \begin{algorithmic}[1]
    \Statex \textbf{Input:} Groups $G_1,\dots,G_m$, oracles $U_{f_1},\dots,U_{f_m}$
    \Statex \textbf{Output:} Subgroup $A = K^{\perp\perp} = K$
    
    \For{\textbf{each node} $i=1$ \textbf{to} $m$ \textbf{in parallel}}
      \State $K_i^\perp \gets \textsc{EDKL}(G_i, U_{f_i})$; \Comment{Compute local orthogonal subgroup}
      \State Transmit $K_i^\perp$ to the central node via \textsc{classical communication};
    \EndFor
    
    \Statex \textbf{Central node processing:}
    \State Aggregate $K^\perp \gets K_1^\perp \oplus K_2^\perp \oplus \cdots \oplus K_m^\perp$;
    \State Solve linear congruence system to obtain $A = K^{\perp\perp}$;
    
    \State \Return $A = K^{\perp\perp} = K$
  \end{algorithmic}
\end{algorithm}

\begin{thm}\label{thm14}
In Algorithm~\ref{EDK}, the following holds:
\begin{enumerate}[label=\textup{(\roman*)}, itemsep=0pt,align=left,labelwidth=1.5em]
    \item After performing $\mathrm{len}(G_i)-\mathrm{len}(K_i)$ iterations at each node $i~(1\le i\le m)$, followed by classical communication aggregation of the outcomes, we can exactly obtain $A=K$;
    \item Quantum query complexity of Algorithm~\ref{EDK} is $\mathop{\max}\limits_{1\leq i\leq m}3\left( \mathrm{len}(G_i) - \mathrm{len}(K_i) \right)$.
\end{enumerate}

\end{thm}

\begin{proof}
(i)
By Proposition~\ref{thm10}, in Algorithm~\ref{EDKL}, at the $i$-th node $(1\le i\le m)$, after performing $\mathrm{len}(G_i)-\mathrm{len}(K_i)$ iterations, we can exactly obtain $A_i=K_i^\perp$.
Then we have $\Pr(A=K)=\Pr\left(\displaystyle\bigcap_{i=1}^m (A_i=K_i^\perp)\right)=\displaystyle\prod\limits_{i=1}^{m}\Pr(A_i=K_i^\perp)=1$.

(ii) Each iteration in Algorithm~\ref{EDKL} needs to query oracle $U_{f_i}$ 3 times, once from operator $\mathcal{A}$ and twice from operator $Q_j$. 

Hence, at node \(i\) (\(1 \le i \le m\)), quantum queries to oracle \(U_{f_i}\) is  $3(\mathrm{len}(G_i)-\mathrm{len}(K_i))$. Due to the parallelism across nodes, the quantum query complexity of Algorithm~\ref{EDK} is dominated by the maximum cost among all nodes, i.e., $\mathop{\max}\limits_{1 \leq i \leq m} 3 \left( \mathrm{len}(G_i) - \mathrm{len}(K_i) \right)$.
   
Overall, the asymptotic complexity of Algorithm~\ref{EDK} is $\mathcal{O}\left( \mathop{\max}\limits_{1 \leq i \leq m} \left( \mathrm{len}(G_i) - \mathrm{len}(K_i) \right) \right)$.
\end{proof}

\begin{remark}
Even if we disregard any benefits from parallelism and consider the total resource consumption in the distributed scenario, the sum of quantum queries across all nodes equals that in the centralized scenario: 
\[
\sum_{i=1}^{m} 3\left(\mathrm{len}(G_i) - \mathrm{len}(K_i) \right)=3\left( \mathrm{len}(G) - \mathrm{len}(K) \right).
\]

This equality demonstrates that Algorithm~\ref{EDK} not only reduces local query complexity but also maintains global computational costs, distinguishing it from the existing distributed quantum algorithms for Simon's problem~\cite{Tan2022DQCSimon} and generalized Simon's problem~\cite{li2024exact}.
\end{remark}

\begin{remark}
 We compare distributed quantum Algorithm~\ref{EDK} against centralized Algorithms~\ref{algorithm1} and~\ref{algorithm3} in Table~\ref{fenbushiliangzisuanfabiao}.
Consider the group $G =\mathop{\oplus}\limits_{i=1}^m G_i=\mathop{\oplus}\limits_{i=1}^m \mathop{\oplus}\limits_{j = 1}^{r_i} \mathbb{Z}_{p_i^{\alpha_{ij}}}$, where $p_1,\dots,p_m$ are distinct primes and $\alpha_{ij} \ge 1$. Algorithm~\ref{EDK} employs $m$ nodes and significantly reduces the resource requirements per node. In terms of qudit number, each node $i$ ($1 \leq i \leq m$) needs $r_i+\sum\limits_{i=1}^{m} r_i$ qudits, much lower than that of $2\sum\limits_{i=1}^{m}r_i$ qudits in Algorithms~\ref{algorithm1} and~\ref{algorithm3}, where $r_i$ is the qudit number to encode group $G_i$ and $\sum\limits_{i=1}^{m} r_i$ is the qudit number to encode group $G$ and set $S$ (see Definition~\ref{HSPdingyi}). For quantum queries, the per‑node complexity of Algorithm~\ref{EDK} is $\mathop{\max}\limits_{1 \leq i \leq m} 3 \left( \mathrm{len}(G_i) - \mathrm{len}(K_i) \right)$, much lower than the centralized complexity $3\bigl(\operatorname{len}(G) - \operatorname{len}(K)\bigr)$ of Algorithm~\ref{algorithm3}. By reducing qudits per node, our distributed quantum Algorithm~\ref{EDK} needs shallower oracle circuits, enhancing noise resistance for NISQ-era implementation.

Table~\ref{fenbushisimonbijiaobiao} shows the comparison of Algorithm~\ref{EDK} with other distributed quantum algorithm for Simon's problem in~\cite{Tan2022DQCSimon} and generalized Simon's problem in~\cite{li2024exact}. Algorithm~\ref{EDK} is the first distributed quantum algorithm for finite AHSP without quantum communication, whereas distributed algorithms in~\cite{Tan2022DQCSimon,li2024exact} require $\mathcal{O}((n-t)(2^{t} - 1)(n-t+l))$ quantum communication. Moreover, Algorithm~\ref{EDK}'s node count ($m$) is determined by the number of prime factors of $|G|$, reducing from $2^t$ (exponential) nodes in distributed algorithms in~\cite{Tan2022DQCSimon, li2024exact}, which marks a substantial improvement in resource efficiency.

\end{remark}

\subsection{Extensions to non-Abelian groups}\label{Extensions to non-Abelian groups}

By Theorem~\ref{Subgroup Decomposition via Direct Products}, our LOCC distributed method can be generalized to a class of non-Abelian groups. For example, consider a non-Abelian group $G$ of order 168. If analysis of its multiplication table reveals that $G \cong G_1 \times G_2 = (\mathbb{Z}_3 \ltimes \mathbb{Z}_7) \times (\mathbb{Z}_2 \wr \mathbb{Z}_2)$, where $\ltimes$ denotes the semidirect product and $\wr$ the wreath product. Then since $|\mathbb{Z}_3 \ltimes \mathbb{Z}_7| = 21$ and $|\mathbb{Z}_2 \wr \mathbb{Z}_2| = 8$ are coprime, Theorem~\ref{Subgroup Decomposition via Direct Products} enables a distributed hidden subgroup search in $G_1$ and $G_2$ independently. The hidden subgroup in $G_1$ can be found by \cite{moore2004power}, while the hidden subgroup in $G_2$ found by~\cite{roetteler1998polynomial}.

\section{Parallel exact classical algorithm for finite AHSP}\label{Sec6}

The current state-of-art deterministic classical algorithms for AHSP exhibit the following query complexities: 
deterministic algorithm in~\cite{nayak2022deterministic} uses $\mathcal{O}(\log |K|\cdot\sqrt{\frac{|G|}{|K|}})$ classical queries, while deterministic algorithm in~\cite{ye2022deterministic} needs $\mathcal{O}(\sqrt{\frac{|G|}{|K|}\log|K|}+\log|K|)$ classical queries. The latter achieves a better asymptotic query complexity.

However, neither of these two classical algorithms~\cite{nayak2022deterministic, ye2022deterministic} utilizes the structure of Sylow \( p \)-subgroups. Consequently, both can be distributed according to Theorem~\ref{External Direct Sum Decomposition of Subgroups}. Moreover, even in a centralized implementation, these algorithms can be further optimized.
We can rewrite group $G\cong\mathbb{Z}_{p_1^{\alpha_1}} \oplus \mathbb{Z}_{p_2^{\alpha_2}} \oplus \cdots \oplus \mathbb{Z}_{p_k^{\alpha_k}}$, where the $p_i$ are primes and $\alpha_i \geq 1$, by regrouping components sharing the same prime into subgroups:
\[
G \cong \mathop{\oplus}\limits_{i=1}^m G_i, \quad \text{with} \quad G_i = \mathop{\oplus}\limits_{j = 1}^{r_i} \mathbb{Z}_{p_i^{\alpha_{ij}}},
\]
where $p_1,\cdots,p_m$ are now distinct primes, $\alpha_{ij} \geq 1$, and $\sum\limits_{i=1}^m r_i = k$.

Then we have Algorithm~\ref{Exact parallel classical algorithm for finding $K_i$} for finding $K_i$. In the parallel classical algorithm, $m$ nodes are required. By part (i) of Theorem~\ref{External Direct Sum Decomposition of Subgroups}, $K = K_1 \oplus K_2 \oplus \cdots \oplus K_m$. 
This decomposition enables distributed computation: each node \(i\) (\(1 \le i \le m\)) locally finds its subgroup \(K_i\) within the Abelian \(p_i\)-group \(G_i\), transmits the result to the central node via classical communication, and the central node aggregates all $K_i$ to reconstruct $K$. The sole distinction between Algorithm~\ref{Exact parallel classical algorithm for finding $K_i$} and Algorithm 4 in~\cite{ye2022deterministic} is substituting $(G, K, k,f)$ with $(G_i, K_i, r_i, f_i)$, where sub-function $f_i$ is provided in Definition~\ref{zihanshudingyi}.

Our algorithm is fully parallelized and the whole parallel classical algorithm is formally presented in Algorithm~\ref{EDCK}. Then we give Theorem~\ref{thm18} which guarantees the correctness of Algorithm~\ref{EDCK}.

\begin{algorithm}[h]
\caption{Exact parallel classical algorithm for finding $K_i$}
\label{Exact parallel classical algorithm for finding $K_i$}

\begin{algorithmic}[1]
\Statex Input: $G_i=\mathop{\oplus}\limits_{j = 1}^{r_i} \mathbb{Z}_{p_i^{\alpha_{ij}}}$\quad ($\alpha_{ij}\geq 1$)
\Statex Output: The hidden subgroup $K_i$
\Procedure{EDCKL}{}
    \State $V = W_1 = W_2 = K_0=O_{r_i},\ r = 0$; \Comment{Initialize variables}
    \For{$l = 1 \to r_i$ }
        \State $t_l = \alpha_{il}$; \Comment{Initialize loop bound}
        \State Query all not queried elements in $W_1$;
        \For{$j = 0 \to \alpha_{il} - 1$}
            \State Query elements in $W_2 + w_l^j$;
            \If{there exist $x \in W_1,\ y \in W_2 + w_l^j$ such that $f_i(x) = f_i(y)$ }
                \State $K_l \gets K_{l-1} + \langle y - x \rangle,\ t_l = j$; \Comment{Update subgroup}
                \If{$j = 0$} $r=r+1$;
                \EndIf
                \State \textbf{break};
            \EndIf
        \EndFor
        \If{$t_l = \alpha_{il}$} $K_l \gets K_{l-1}$; \Comment{Trivial subgroup case}
        \EndIf
        \State $V \gets V + V_l^{t_l}$; \Comment{Update search subgroup}
        \State $W_1, W_2 \gets \text{findPair}(V, \max\{1, r\})$;
    \EndFor
    \State \Return $K_{r_i}(=K_i)$
\EndProcedure
\end{algorithmic}
\end{algorithm}

For Algorithm~\ref{Exact parallel classical algorithm for finding $K_i$}:
\begin{itemize}
    \item In line 7, $w_l^j =( \underbrace{0,\ldots,0}_{l-1},p_i^j,\underbrace{0,\ldots,0}_{r_i-l})$.
    \item In line 17, $V_l^{t_l} = O_{l-1} \times \mathbb{Z}_{p_i^{t_l}} \times O_{r_i-l}$.
    \item In line 18, $\text{findPair}$ is a subroutine that doesn't make any query in~\cite{ye2022deterministic}.
\end{itemize}

\begin{algorithm}[t]
  \caption{Exact parallel classical algorithm for finding the entire subgroup $K$}\label{EDCK}
  \begin{algorithmic}[1]
    \Procedure{EDCK~}{groups $G_1,\dots,G_m$}\strut
      \For{\textbf{each node} $i=1$ \textbf{to} $m$ \textbf{in parallel}}
        \State Execute \textsc{EDCKL} at node $i$ to compute $K_i$;
        \State Transmit $K_i$ to the central node via \textsc{classical communication};
      \EndFor
      \State \textbf{Central node:} 
       Aggregate  $A\gets K_1 \oplus K_2 \oplus \cdots \oplus K_m$;
      \State \Return $A=K$
    \EndProcedure \strut
  \end{algorithmic}
\end{algorithm}

\begin{thm}\label{thm18}
In Algorithm~\ref{EDCK} (EDCK), the following holds:
\begin{enumerate}[label=\textup{(\roman*)}, itemsep=0pt,align=left,labelwidth=1.5em]
    \item After querying $\mathcal{O}(\sqrt{\dfrac{|G_i|}{|K_i|}\log|K_i|}+\log|K_i|)$ times at each node $i(1\le i\le m)$, followed by classical communication aggregation of the outcomes, we can exactly obtain $A=K$;
    \item Exact query complexity of Algorithm~\ref{EDCK} is $\mathcal{O}\left(\mathop{\max}\limits_{1\leq i\leq m}(\sqrt{\dfrac{|G_i|}{|K_i|}\log|K_i|}+\log|K_i|)\right)$. 
\end{enumerate}

\end{thm}

\begin{proof}
(i)
The sole distinction between Algorithm~\ref{Exact parallel classical algorithm for finding $K_i$} and Algorithm 4 in~\cite{ye2022deterministic} lies in replacing $(G, K, k,f)$ with $(G_i, K_i, r_i,f_i)$. Thus, we can exactly 
obtain $K_i$ at the $i$-th node$(1\le i\le m)$. As $K = K_1 \oplus K_2 \oplus\cdots\oplus K_m$, we can exactly obtain $A=K$ in the central node.

(ii)
     At node \(i\)(\(1 \le i \le m\)), query complexity is  
    \[
   \mathcal{O}(\sqrt{\dfrac{|G_i|}{|K_i|}\log|K_i|}+\log|K_i|).
    \]
    Due to the parallelism across nodes, the exact query complexity of Algorithm~\ref{EDCK} is dominated by the maximum cost among all nodes, i.e.,
    \[
 \mathcal{O}\left(\mathop{\max}\limits_{1\leq i\leq m}(\sqrt{\dfrac{|G_i|}{|K_i|}\log|K_i|}+\log|K_i|)\right). 
    \]

\end{proof}

\begin{remark}
Even if we disregard any benefits from parallelism and consider only the total resource consumption in the distributed scenario, the sum of queries across all nodes does not exceed that in the original algorithm:

\begin{align}
   \sum\limits_{i=1}^{m}\left(\sqrt{\dfrac{|G_i|}{|K_i|}\log|K_i|}+\log|K_i|\right)
   &= \sum\limits_{i=1}^{m}\sqrt{\dfrac{|G_i|}{|K_i|}}\cdot\sqrt{\log|K_i|}+\log|K| 
   \quad \bigl(\text{since } \prod_{i=1}^{m}|K_i|=|K|\bigr) \nonumber\\
&\leq \sqrt{ \left( \sum_{i=1}^m \frac{|G_i|}{|K_i|} \right)} \cdot\sqrt{\left( \sum_{i=1}^m \log |K_i|\right)}+\log|K| 
\quad \text{(Cauchy–Schwarz inequality)} \nonumber\\
&\leq \sqrt{ \left( \prod_{i=1}^{m}\frac{|G_i|}{|K_i|} \right)} \cdot\sqrt{\left( \sum_{i=1}^m \log |K_i|\right)}+\log|K|\label{shizi155} \\
&= \sqrt{\frac{|G|}{|K|} \cdot \log|K|}+\log|K|,
\quad \bigl(\text{since } \prod_{i=1}^{m}|G_i|=|G| \text{ and } \prod_{i=1}^{m}|K_i|=|K|\bigr)\nonumber
\end{align}

\begin{itemize}
\item  ``$\leq$'' in Eq.~(\ref{shizi155}) holds under the condition that $\displaystyle\prod_{i=1}^{m}\frac{|G_i|}{|K_i|} \geq \sum_{i=1}^m \frac{|G_i|}{|K_i|}$, which holds when $|G_i|/|K_i|\geq 2$ for all $i$ (i.e., $K_i \neq G_i$). This is a sufficient but not necessary condition.
\end{itemize}

In summary, we have established the inequality:
\[
\sum_{i=1}^{m} \left( \sqrt{\frac{|G_i|}{|K_i|} \log |K_i|} + \log |K_i| \right) \leq \sqrt{\frac{|G|}{|K|} \cdot \log |K|} + \log |K|
\]
under the conditions $K_i \neq G_i$. This implies the asymptotic containment:
\[
\mathcal{O}\left( \sum_{i=1}^{m} \left( \sqrt{\frac{|G_i|}{|K_i|} \log |K_i|} + \log |K_i| \right) \right) \subseteq \mathcal{O}\left( \sqrt{\frac{|G|}{|K|} \cdot \log |K|} + \log |K| \right).
\]

This demonstrates that the complexity reduction of our parallel classical algorithm is fundamental under the \textbf{sufficient (not necessary) condition} that \( K_i \subsetneqq  G_i\quad (i = 1, 2, \ldots, m)\), which is easy to satisfy in practical scenario.

\end{remark}

\section{Conclusion and future works}\label{Sec7}

In this paper, we first present the iteration count for achieving success probability $1-\epsilon$ in the standard quantum algorithm for finite AHSP to be either $\mathrm{rank}(G) + \lceil \log_2 (2/\epsilon) \rceil$ or $\mathrm{len}(G)+ \lceil \log_2 (1/\epsilon) \rceil$. The former offers an exponential improvement in $\epsilon$-dependence over the prior bound $\lfloor 4/\epsilon \rfloor \mathrm{rank}(G)$~\cite{koiran2007quantum}, while the latter improves upon $\lceil \log_2 |G| + \log_2 (1/\epsilon) + 2 \rceil$~\cite{lomont2004hidden}. Then we present an exact quantum algorithm for finite AHSP, when the order $|K|$ of subgroup $K$ is known. Our algorithm is more concise than the previous exact algorithm for finite AHSP~\cite{imran2022exact}, and applies to any finite Abelian group $G$.

Leveraging Chinese Remainder Theorem, we propose a distributed exact quantum algorithm for finite AHSP, which is a pure LOCC model. Compared to the centralized finite AHSP, our distributed algorithm requires fewer qudits and has lower query complexity. We also give a parallel exact classical algorithm for finite AHSP, which has lower query complexity, even disregarding any benefits from parallelism, the sum of queries at all nodes is not greater than that in centralized algorithm under a condition that is easily satisfied.

In Sec.~\ref{Extensions to non-Abelian groups}, we demonstrate that our LOCC distributed method can be extended to a class of non-Abelian groups, provided that the orders of all factors in the direct product decomposition are pairwise coprime. Beyond direct product, there exists other methods of constructing larger groups from smaller ones, such as semidirect products and group extensions, raising several open questions about \emph{alternative} distributed LOCC frameworks that merit investigation:

\begin{itemize}

    \item \textbf{Semidirect product groups:} For non-Abelian groups $G = G_1 \rtimes_\varphi G_2$, does a distributed LOCC framework exist? 

    \item \textbf{General group extensions:} For non-Abelian groups fitting into a group extension $1 \to G_1 \to G \to G_2 \to 1$ (short exact sequence), does a distributed LOCC framework exist?
    
    \item \textbf{Infinite groups:} For the HSP on infinite groups (where our current framework applies only to finite groups), does a distributed LOCC framework exist?
\end{itemize}

\section*{Acknowledgements}
This work is supported in part by the National Natural Science Foundation of China (Nos. 61876195, 11801579), and the Natural Science Foundation of Guangdong Province of China (No. 2017B030311011).

\bibliographystyle{unsrt}
\bibliography{ref}

\end{document}